\documentclass[a4paper, 11pt]{article}
\pdfoutput=1

\usepackage{jcappub}
\usepackage{graphicx}
\usepackage{booktabs}

\newcommand{\Er}{E_{\rm{R}}}
\newcommand{\vmin}{v_{\rm{min}}}
\newcommand{\mDM}{m_{\rm{DM}}}
\newcommand{\rhoDM}{\rho_{\rm{DM}}}
\newcommand{\Sone}{\mathrm{S1}}
\newcommand{\Stwo}{\mathrm{S2}}
\newcommand{\Stwob}{\mathrm{S2_b}}
\newcommand{\logSbS}{\log_{10}(\mathrm{S2_b}/\mathrm{S1})}

\newcommand{\nquanta}{n_{\mathrm{quanta}}}
\newcommand{\nion}{n_{\mathrm{i}}}
\newcommand{\nex}{n_{\mathrm{ex}}}
\newcommand{\Nquanta}{N_{\mathrm{quanta}}}
\newcommand{\Nion}{N_{\mathrm{i}}}
\newcommand{\Nex}{N_{\mathrm{ex}}}
\newcommand{\nele}{n_e}
\newcommand{\ngam}{n_{\gamma}}

\newcommand{\xea}{^{129}\mathrm{Xe}}
\newcommand{\xeb}{^{131}\mathrm{Xe}}
\newcommand{\xec}{^{136}\mathrm{Xe}}

\newcommand{\kr}{^{85}\mathrm{Kr}}
\newcommand{\rn}{^{222}\mathrm{Rn}}

\newcommand{\muDM}{\mu_{\mathrm{DM}}}
\newcommand{\muBG}{\mu_{\mathrm{BG}}}
\newcommand{\muBGj}{\mu_{\mathrm{BG}j}}
\newcommand{\muBGk}{\mu_{\mathrm{BG}k}}
\newcommand{\fDM}{f_{\mathrm{DM}}}
\newcommand{\fBG}{f_{\mathrm{BG}}}
\newcommand{\fBGj}{f_{\mathrm{BG}j}}

\title{
Prospects for dark matter detection with inelastic transitions of xenon
}
\author{Christopher McCabe}
\affiliation{GRAPPA Centre for Excellence,\\Institute for Theoretical Physics Amsterdam (ITFA),\\
 University of Amsterdam, Science Park 904, the Netherlands}
 \emailAdd{c.mccabe@uva.nl}

\abstract{
Dark matter can scatter and excite a nucleus to a low-lying excitation in a direct detection experiment. This signature is distinct from the canonical elastic scattering signal because the inelastic signal also contains the energy deposited from the subsequent prompt de-excitation of the nucleus. A measurement of the elastic and inelastic signal will allow a single experiment to distinguish between a spin-independent and spin-dependent interaction. For the first time, we characterise the inelastic signal for two-phase xenon detectors in which dark matter inelastically scatters off the~$\xea$ or~$\xeb$ isotope. We do this by implementing a realistic simulation of a typical tonne-scale two-phase xenon detector and by carefully estimating the relevant background signals. With our detector simulation, we explore whether the inelastic signal from the axial-vector interaction is detectable with upcoming tonne-scale detectors. We find that two-phase detectors allow for some discrimination between signal and background so that it is possible to detect dark matter that inelastically scatters off either the~$\xea$ or~$\xeb$ isotope for dark matter particles that are heavier than approximately $100$~GeV. If, after two years of data, the XENON1T search for elastic scattering nuclei finds no evidence for dark matter, the possibility of ever detecting an inelastic signal from the axial-vector interaction will be almost entirely excluded.
 }

\begin{document}
\maketitle
\flushbottom

\section{Introduction}
The properties of particle dark matter remain unknown. Searches with direct detection experiments are one of the most promising ways of detecting dark matter through an interaction other than gravity. A positive detection is expected to yield information on the particle mass, the cross-section and information on the form of the interaction~\cite{Peter:2013aha,Catena:2014epa}. Although there has not yet been a conclusive detection~\cite{Bernabei:2013xsa, Pradler:2012qt, Aalseth:2014eft, Davis:2014bla, Angloher:2011uu, Brown:2011dp,Kuzniak:2012zm,Agnese:2013rvf}, direct detection experiments have demonstrated a remarkable record of increasing their sensitivity by an order of magnitude approximately every three years and this increase is expected to continue over the next decade~\cite{Cushman:2013zza}. Two-phase xenon experiments have proven to be particularly sensitive and we are approaching the tonne-scale era with the LUX~\cite{Akerib:2012ys} and XENON1T~\cite{Aprile:2012zx} experiments, and funding has been secured for the approximately five-tonne successor experiments, LZ~\cite{Malling:2011va,Akerib:2015cja} and XENONnT~\cite{Aprile:2014zvw,Aprile:2015uzo}. There is also a longer-term proposal for DARWIN~\cite{Baudis:2012bc}, an even larger $\sim20$-tonne experiment whose aim is to explore all of the dark matter parameter space not limited by neutrino backgrounds~\cite{Billard:2013qya}.

Multi-tonne xenon experiments bring new opportunities to search for rare signals. This is for two reasons. Firstly, the larger target mass means that there are more xenon nuclei for the dark matter to scatter off and secondly, larger experiments allow for backgrounds to be significantly reduced, even down to the irreducible background from coherent neutrino scattering. This is because more of the liquid xenon can be used to self-shield the fiducial volume where dark matter signals are searched for.

\begin{figure}[t!]
\centering
\includegraphics[width=0.43\columnwidth]{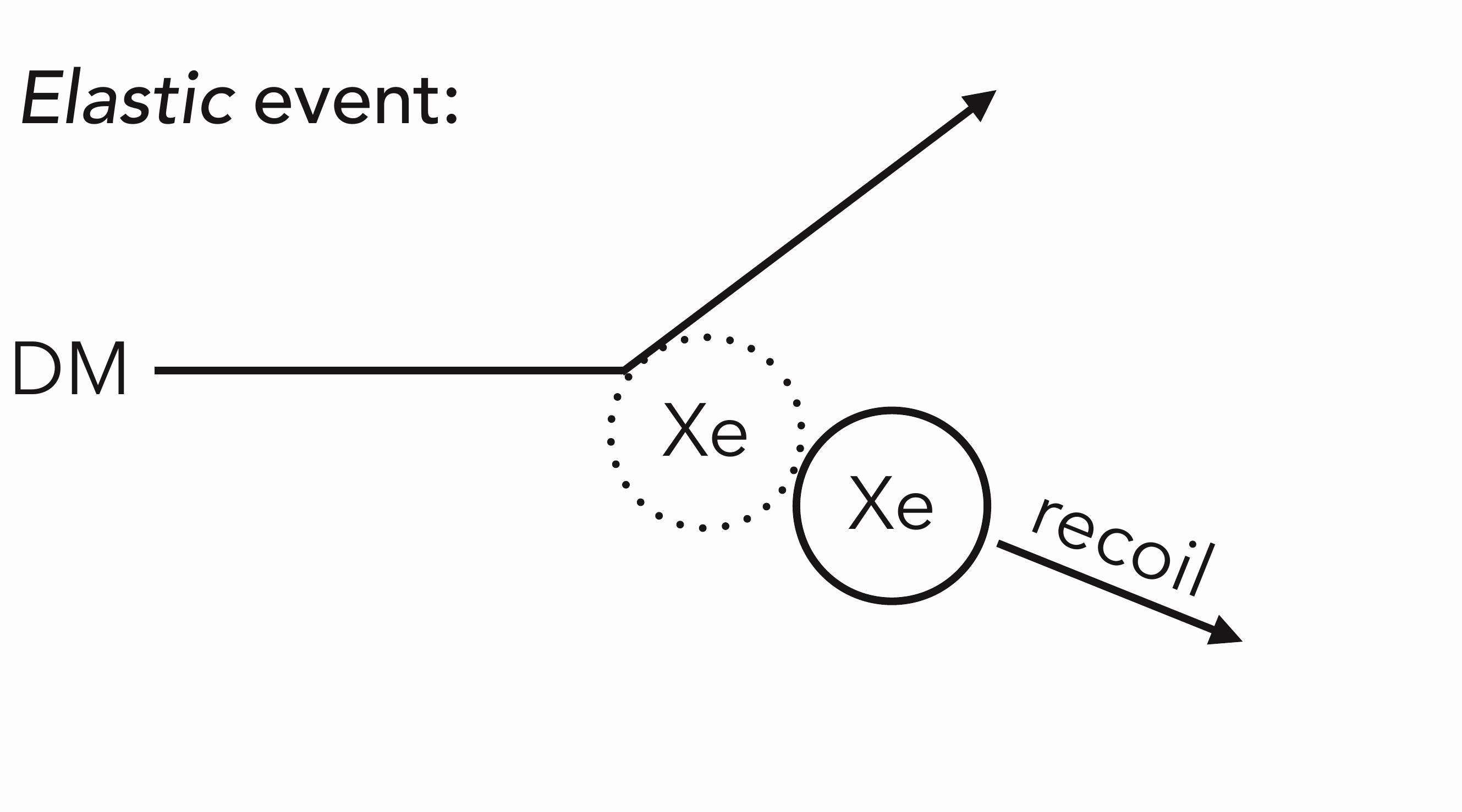} 
\includegraphics[width=0.43\columnwidth]{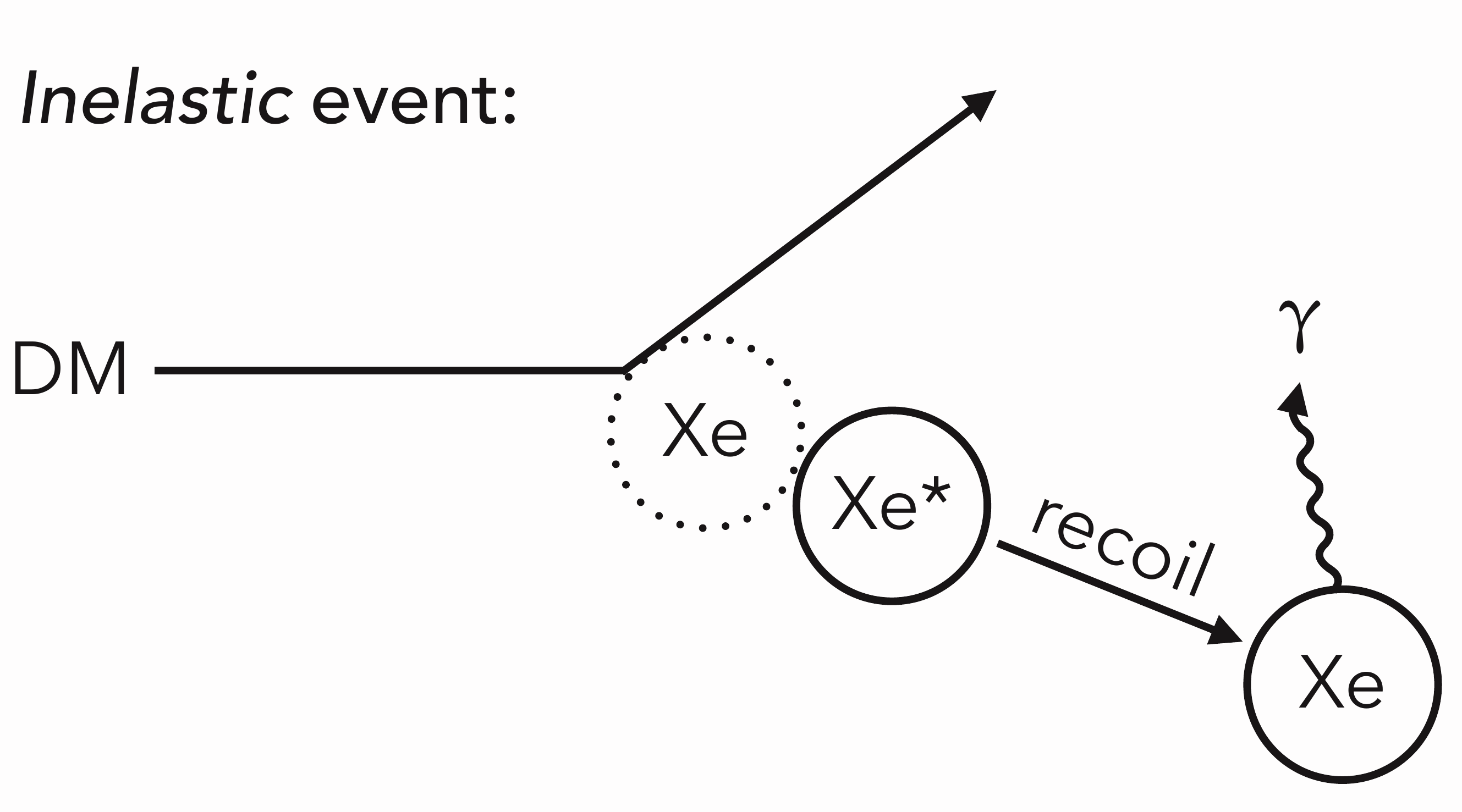} 
 \caption{The left and right diagrams depict two dark matter signals at a direct detection experiment. The left diagram shows the canonical elastic scattering process where the dark matter simply causes the nucleus to recoil; an experiment measures the number of events and the nuclear recoil energy. The right diagram depicts the inelastic scattering process. In this case, the dark matter excites the xenon isotope which then promptly decays emitting a photon. For the~$\xea$ and~$\xeb$ isotopes of xenon, the photon/excitation energies are $39.6~\mathrm{keV}$ and $80.2~\mathrm{keV}$ respectively. We assume that the photon mean free path is sufficiently short that the experiment measures the recoil of the nucleus at the same time as the prompt de-excitation photon.
 \label{fig:scattering}}
 \end{figure}

The canonical search with direct detection experiments is the elastic scattering process depicted in the left diagram of figure~\ref{fig:scattering}. The interaction with the dark matter particle causes the xenon nucleus to recoil with an energy typically in the range $1$\,--\,$100~\mathrm{keV}$. Since some nuclear isotopes have excitations in this energy range, it was long ago realised that these nuclear excitations could also play a role in the detection of dark matter~\cite{Goodman:1984dc, Ellis:1988nb}. In this case, some part of the energy transferred from the dark matter particle causes the excitation of the nucleus while the other part causes the nucleus to recoil. The excited nucleus then decays emitting a photon. This process is depicted in the right panel of figure~\ref{fig:scattering}. For experiments with xenon, there are two isotopes of interest, $\xea$ and $\xeb$, which make up $26.4\%$ and $21.2\%$ of natural xenon and have an excitation energy and lifetime of $39.6~\mathrm{keV}$ and $80.2~\mathrm{keV}$, and $0.97~\mathrm{ns}$ and $0.48~\mathrm{ns}$ respectively. In this process the recoil energy of the nucleus and the energy from the prompt de-excitation of the nuclear isotope are measured. The experimental resolution of xenon detectors is $\sim10~\mathrm{ns}$~\cite{Aprile:2011dd} so the short lifetimes mean that the recoil and de-excitation cannot be separately resolved. However, the mean free path of the de-excitation photon is $\mathcal{O}(1)~\mathrm{mm}$~\cite{Malling:2014wza}, comparable to the spatial resolution of xenon detectors~\cite{Aprile:2011dd}, so a dedicated pulse-shape analysis may partially resolve the nuclear recoil and the photon energy deposition for a fraction of the events. We leave an analysis of the pulse-shape for the future and here make the assumption that the detector cannot resolve the recoil energy and photon energy i.e.\ it is only the total energy that is measured. 

Nuclear structure functions are required in order to accurately predict the cross-section for dark matter to excite the nucleus. It is only recently that precision shell-model calculations of the structure functions for xenon isotopes have become available~\cite{Klos:2013rwa,Baudis:2013bba,Vietze:2014vsa} (see also~\cite{Toivanen:2008zz,Toivanen:2009zza} for earlier calculations). The contribution of different nucleons to inelastic scattering does not add coherently so this process does not benefit from the nucleon-number--squared $(\sim 10^4$) enhancement of elastic spin-independent interactions~\cite{Vietze:2014vsa}. The current absence of any elastic signal means that for these interactions, experiments would have to improve their sensitivity by at least this factor to begin to see the inelastic signal. Such a large sensitivity gain is unlikely to be achieved in the foreseeable future. In contrast, the structure functions for elastic and inelastic processes are more comparable for the axial-vector interaction, with the elastic structure function being only around $10$ times larger~\cite{Baudis:2013bba}. This is because elastic scattering for the axial-vector interaction is spin-dependent and also does not have the the nucleon-number--squared enhancement~\cite{Klos:2013rwa}. The initial discovery of dark matter will not be made with the inelastic process for the axial-vector interaction (because of the additional suppression of the inelastic rate from the structure function and the scattering kinematics). However, a detection of the inelastic signal would provide additional information to complement the elastic scattering signal. As a trivial example, while the elastic signal may come from either a spin-independent or spin-dependent interaction, the inelastic signal will only be detectable for a spin-dependent interaction so it detection would strongly point to a spin-dependent interaction. Further implications of measuring the inelastic signal are left for a future paper.

A number of {\it single-phase} xenon experiments have searched for the $39.6~\mathrm{keV}$ de-excitation from the $\xea$ isotope~\cite{Belli:1991mx,Bernabei:2000qn,Uchida:2014cnn} (see also~\cite{Vergados:2013raa}). However these experiments generally set weaker limits than {\it two-phase} xenon experiments because they do not have the same ability as two-phase experiments to discriminate between signal and background processes. No search or sensitivity study has been carried out for a {\it two-phase} xenon detector. This is the aim of this paper: to characterise the inelastic scattering signal for a {\it two-phase} xenon detector, quantify the sensitivity of upcoming tonne-scale experiments to this inelastic process and assess whether a future detection can be made. 
 
Our paper is structured as follows. In section~\ref{sec:model} we recap the basic principles of dark matter scattering and describe how we model the elastic and inelastic signals in two benchmark xenon detectors. In section~\ref{sec:backgrounds} we discuss the main backgrounds and calculate both the signal and background distributions in terms of the parameters that a xenon detector measures. Section~\ref{sec:discovery} describes our frequentist method for calculating the sensitivity of upcoming tonne-scale experiments while section~\ref{sec:results} contains our main results. We end with a discussion of interesting follow-up studies and our conclusions in section~\ref{sec:con}. A number of short appendices gather the formulae that we use for the generation of photons and electrons for nuclear and electronic interactions, a check of the statistical method that we employ, an explicit demonstration that the LUX neutron-only limits are generally stronger than the PICO proton-only limits and finally, a check of our results under an alternative dark matter halo model.

\section{Modelling elastic and inelastic recoils of xenon}
\label{sec:model}

In this section we first review the usual formalism for elastic scattering of dark matter with a xenon nucleus in terms of the recoil energy of the nucleus. We show that this is easily extended to the case of inelastic scattering. Xenon detectors do not directly measure the energy but rather the scintillation light. We describe our modelling of the generation and detection of the scintillation light, which is based on the NEST formalism~\cite{Szydagis:2011tk,Szydagis:2013sih,Mock:2013ila,Lenardo:2014cva}. We then describe the properties of present and upcoming tonne-scale direct detection experiments and discuss the observable signals and their rate.

\subsection{Scattering rates}

The differential event rate for both elastic and inelastic scattering of dark matter with a xenon nucleus of mass $m_A$ in the detector frame may be written as
\begin{equation}
\label{eq:dRdE}
\frac{d R}{d \Er}=\frac{1}{m_A} \frac{\rhoDM}{\mDM} \int_{\vmin} d^3v \, v f_{\rm{DM}}(\vec{v}+\vec{v}_{\rm{E}}) \frac{d \sigma}{d \Er} \;,
\end{equation}
where $\Er$ is the recoil energy of the xenon nucleus, $\mDM$ is the dark matter mass, $\rhoDM=0.3~\mathrm{GeV}/\mathrm{cm}^3$ is the local dark matter density~\cite{Read:2014qva}, $v$ and $\vec{v}$ are the dark matter speed and velocity, and~$f_{\rm{DM}}(\vec{v})$ is the dark matter velocity distribution in the galactic frame. We assume the isothermal Standard Halo Model so that~$f_{\rm{DM}}(\vec{v})\propto\exp{\left(-v^2/v_0^2 \right)}$ is a Maxwell-Boltzmann distribution in the galactic frame with a hard cut-off at the galactic escape speed~$v_{\rm{esc}}$, for which we assume~$v_{\rm{esc}}=550~\mathrm{km}/\mathrm{s}$~\cite{Piffl:2013mla}. The solar circular speed is by convention taken as~$v_0=220~\mathrm{km}/\mathrm{s}$ and we boost from the galactic frame to the detector rest frame with $\vec{v}_{\rm{E}}=(0,v_0,0)+\vec{v}_{\rm{pec}}+\vec{v}_{\rm{e}}$, where $\vec{v}_{\rm{pec}}=(11.1,12.2,7.3)~\mathrm{km}/\mathrm{s}$~\cite{Schoenrich:2009bx} and we use the expression for $\vec{v}_{\rm{e}}$ from~\cite{McCabe:2013kea}.

The minimum speed to recoil with an energy $\Er$ additionally depends on the excitation energy~$E^*$:
\begin{equation}
\vmin=\sqrt{\frac{m_A \Er}{2 \mu_{A}^2}}+\frac{E^*}{\sqrt{2 m_A \Er}}\;,
\end{equation}
where~$\mu_{A}$ is the nucleus-dark matter reduced mass. The minimum speed is larger for bigger~$E^*$ since part of the kinetic energy of the incoming dark matter particle is required to excite the nucleus. This means that for the same~$\Er$, elastic and inelastic scatter processes probe different parts of~$f_{\rm{DM}}(\vec{v})$~\cite{Baudis:2013bba}.

In this paper we only consider axial-vector interactions of the type
\begin{equation}
\label{eq:A-V}
\mathcal{L}\propto -\bar{\chi}\gamma^{\mu}\gamma^5 \chi \cdot \sum_q A_q \bar{\psi}_q \gamma_{\mu}\gamma^5\psi_q\;,
\end{equation}
where~$\chi$ is the dark matter (here assumed to be a fermion),~$\psi_q$ are the light-quark fields ($q=u,d,s$) and $A_q$ are the (model-dependent) dark matter-quark coupling constants. The total spin-dependent differential cross-section applicable for this operator can generally be written as
\begin{equation}
\frac{d \sigma}{d \Er}=\sum_{A=^{129}\mathrm{Xe},\,^{131}\mathrm{Xe}}\frac{4 \pi}{3}\frac{m_A}{2 \mu_n^2} \frac{\sigma^0_n}{v^2}\frac{f_A}{2 J_A+1}S^n_A(\Er)\;,
\end{equation}
where~$\mu_n$ is the nucleon-dark matter reduced mass, the sum is over the isotopes that have spin, $f_A$ is the fractional abundance of the xenon isotope, $J_A$ is the ground-state spin of the isotope ($J_{129}=1/2$ and $J_{131}=3/2$) and $\sigma^0_n$ is the elastic cross-section to scatter off a point-like neutron in the limit of zero-momentum transfer. The structure factors~$S_{A}(\Er)$ describe how the dark matter interacts with the nucleus and depend on the isotope. We take the central values of the one\,$+$\,two-body expressions from~\cite{Klos:2013rwa} and~\cite{Baudis:2013bba} for elastic and inelastic scattering respectively. In both cases we only consider the neutron structure factors~$S^n_{A}(\Er)$ since the proton structure factors are always at least a factor of 10 smaller.

\begin{figure}[t!]
\centering
\includegraphics[width=0.49\columnwidth]{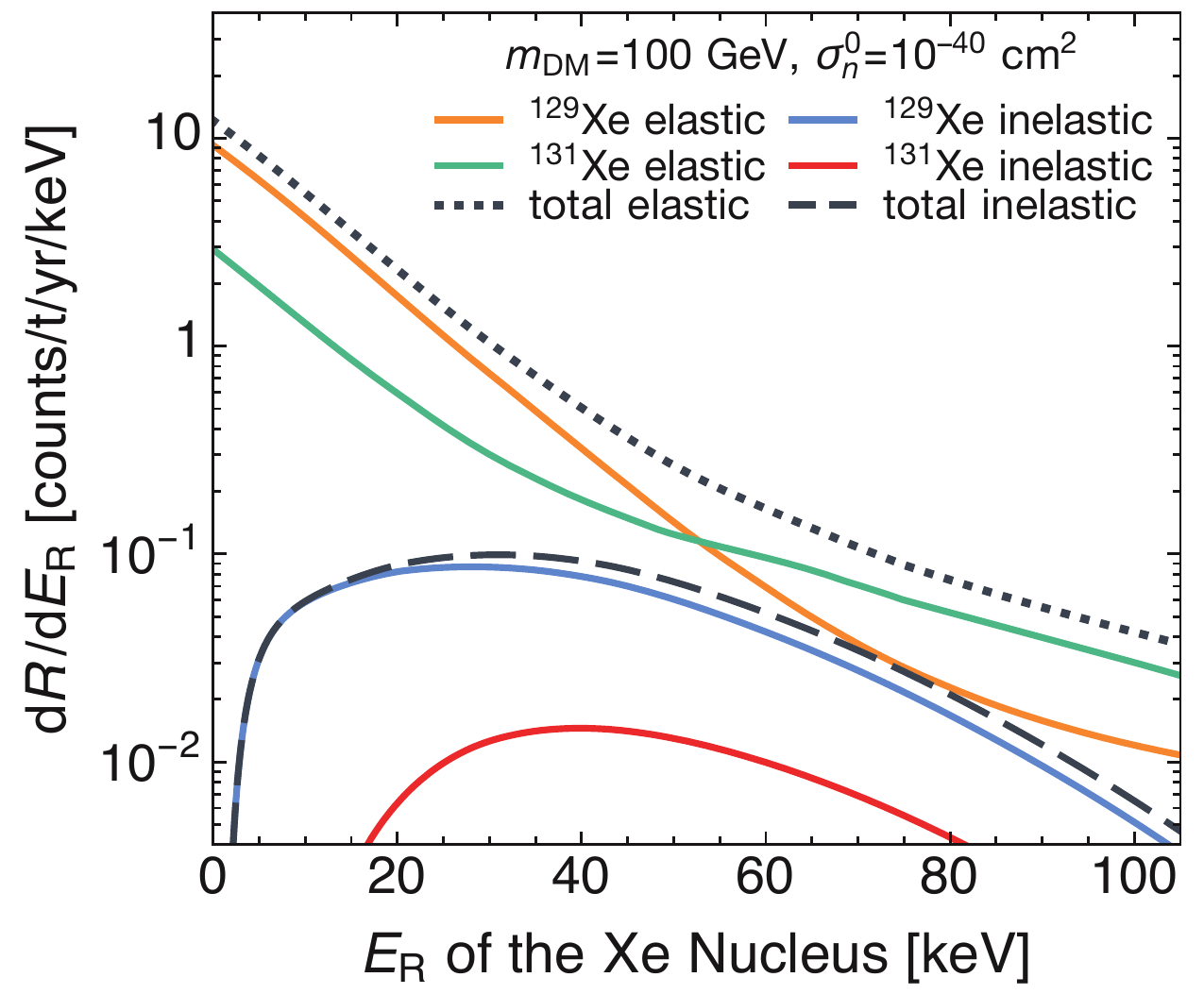} 
\includegraphics[width=0.49\columnwidth]{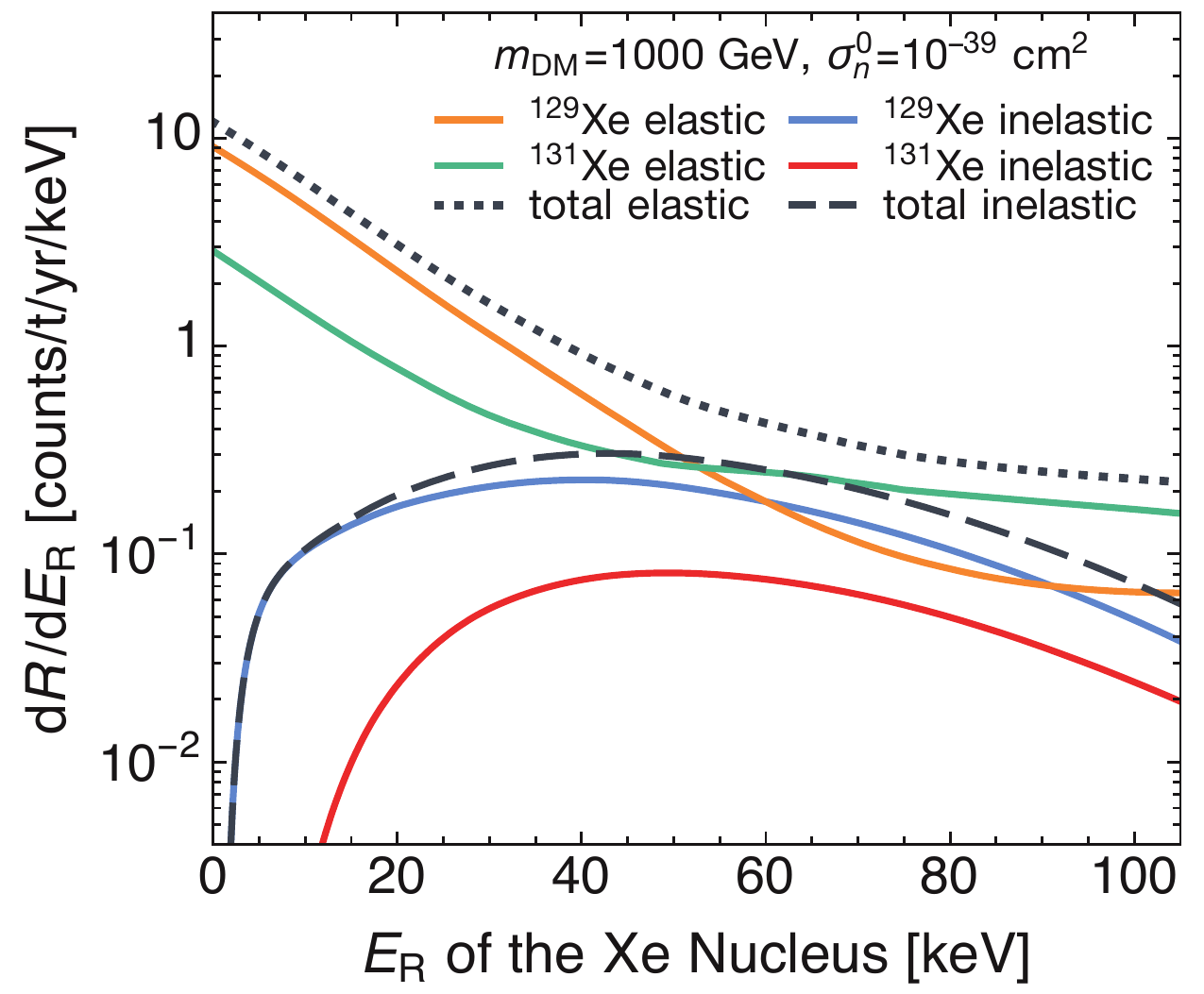} 
 \caption{The left and right panels show the recoil spectra for the elastic and inelastic processes in terms of~$\Er$, the energy of the recoiling nucleus, for two values of the dark matter mass~$\mDM$ and the scattering cross-section~$\sigma^0_n$. The various curves show the individual rates for the xenon isotopes that participate in the scattering for the axial-vector (spin-dependent) interaction, namely~$\xea$ and~$\xeb$, together with the total rate. The elastic rate always dominates implying that the initial discovery of dark matter will always be made with the elastic scattering process. The difference between the elastic and inelastic rates are smaller for heavier particles suggesting that it should be easier to find evidence for the inelastic process with heavier particles. We remind the reader that~$\Er$ is not directly measured in a two-phase xenon detector, rather, it is the scintillation signals~$\Sone$ and~$\Stwo$. \label{fig:spectrum}}
 \end{figure}

The recoil spectra as a function of the xenon nucleus's recoil energy $\Er$ are shown in figure~\ref{fig:spectrum}. The left and right panels show the spectra for $\mDM=100~\mathrm{GeV}$ and $\sigma_n^0=10^{-40}~\mathrm{cm}^2$, and $\mDM=1000~\mathrm{GeV}$ and $\sigma_n^0=10^{-39}~\mathrm{cm}^2$ respectively. The total elastic and inelastic spectrum are shown by the black dotted and black dashed lines respectively. The orange and green lines show the contribution of $\xea$ and $\xeb$ to the elastic spectrum, while the blue and red lines show the contribution of $\xea$ and $\xeb$ to the inelastic spectrum. The elastic spectrum is always larger than the inelastic spectrum with the most noticeable difference at small $\Er$. The inelastic spectrum drops to zero at small $\Er$ because energy and momentum conservation do not allow for the xenon nucleus to be excited while remaining at rest after the dark matter interaction. The larger elastic scattering rate implies that for the axial-vector interaction, a discovery of dark matter will always first be made with the elastic scattering process.

The elastic spectrum is dominated by scattering with $\xea$ at low energy, while scattering with $\xea$ dominates for all energies in the inelastic spectrum. Comparing the left and right panels, we see that the inelastic spectra are closer to the elastic spectra for $\mDM=1000~\mathrm{GeV}$. At low energies and for the mass values shown, the elastic spectra display the characteristic scaling $dR/d\Er\propto \sigma_n^0\, \mDM^{-1}$. This scaling does not continue at higher recoil energies because for smaller masses, it is only the particles in the tail of the Maxwell-Boltzmann distribution that have sufficient kinetic energy to induce higher recoil energies of the xenon nucleus, thus producing an additional suppression in the rate (this is manifested mathematically through a higher value of $\vmin$ for smaller $\mDM$).  The inelastic spectra also do not show the characteristic scaling at any energy for these masses. This is for the same reason as in the elastic case, namely, many more incoming dark matter particles have a larger kinetic energy for higher masses. This is especially noticeable for the~$\xeb$ spectra where there is a factor $\sim6$ difference between the~$\xea$ and~$\xeb$ spectra at~$\mDM=100~\mathrm{GeV}$, while only a factor~$\sim3$ at~$\mDM=1000~\mathrm{GeV}$. This suggests that it should be easier to find evidence for the inelastic process for heavier dark matter particles.

This discussion so far has only considered the recoil energy of the nucleus and has not accounted for the energy deposited by the photon from the de-excitation process. Although the de-excitation will not change the total (integrated) scattering rate, it must be accounted for when modelling the signal that a two-phase xenon detector measures. The next subsections address how we model this.

\subsection{Generating light and charge signals}
\label{sec:gensignal}

Two-phase xenon detectors do not directly measure the energy. Instead, a particle interacting in the fiducial volume of the liquid xenon produces two measurable signals referred to as the S1 and S2 signal.\footnote{We only use the position-corrected S1 and S2 signals (sometimes denoted cS1 and cS2)~\cite{Aprile:2012vw}.} An interaction in the liquid xenon produces ions and excitons which produce photons and electrons. The quantity S1 is a measure of the number of photoelectrons (PE) from the prompt scintillation due to the photons in the liquid xenon. The electrons are drifted in an electric field to the xenon gas phase, where they are extracted and accelerated. These extracted electrons create a secondary scintillation, denoted as S2. Electronic and nuclear events produce different characteristic S1 and S2 signals, which allows two-phase xenon detectors to discriminate between these two event classes. Canonical dark matter interactions result in nuclear events while most background events are electronic events. This ability to discriminate between nuclear and electronic events is one important reason why two-phase xenon detectors have been so successful in constraining the dark matter scattering cross-section. 

For an energy deposition $E$, the expectation values for $\Sone$ and $\Stwo$ can be expressed as $\langle\Sone\rangle=g_1 \langle n_{\gamma}(E) \rangle$ and $\langle\Stwo\rangle=g_2\langle n_e (E) \rangle$ respectively, where the measurement gains $g_1$ and $g_2$ relate the number of produced photons $n_{\gamma}(E)$  and electrons $n_e (E)$ to the expected number of detected PEs.\footnote{The $g_1$ and $g_2$ notation is not used uniformly. It is typically used by the LUX collaboration but different notation exists elsewhere. For instance, refs.~\cite{Baudis:2014naa,Schumann:2015cpa} use $\epsilon$ instead of $g$. Ref.~\cite{Baudis:2014naa} also describes how this notation relates to the description in terms of~$\mathcal{L}_{\rm{eff}}$ and $Q_y$, parameters which some may find more familiar.}  The gain $g_1$ is the probability that a photon produced at the centre of the detector strikes a photomultiplier tube (PMT) and is converted to a~PE. The gain $g_2=\epsilon \times Y$ is the product of the probability of extracting an electron from the liquid to the gas ($\epsilon$) and the amplification factor ($Y$) converting a single ionisation electron to photoelectrons. The S2 signal measured from the bottom PMTs ($\Stwob$) is usually used because the light collection efficiency is more homogeneous on these PMTs~\cite{Aprile:2012vw}. We therefore use $\Stwob$ and assume that it is related to the total $\Stwo$ signal by $\Stwob=0.43\times \Stwo$, as found in XENON100 and LUX~\cite{abrown,Akerib:2013tjd}.  We will discuss realistic values of $g_1$ and $g_2$ in the next subsection and for now, leave them as free parameters in our discussion.

To simulate signal processes observed by a two-phase xenon detector in a realistic fashion, we generate the signal with a Monte Carlo simulation along the lines of ref.~\cite{Dahl:2009nta}. We use the NEST phenomenological model~\cite{Szydagis:2011tk,Szydagis:2013sih,Mock:2013ila,Lenardo:2014cva} to model the average number of photons~$n_{\gamma}(E)$ and electrons~$n_e(E)$ produced by an electronic- or nuclear-type interaction. In addition to their dependence on the energy~$E$,  $n_{\gamma}(E)$ and $n_e(E)$ also depend on the electric drift field applied across the liquid, which varies for different experiments. The specific formulae used in our modelling are given in appendix~\ref{app:meanyields}.  We must include fluctuation effects, which can be divided into two types: intrinsic and detector fluctuations. We discuss the implementation of each in turn beginning with the intrinsic fluctuations.

Our signal generation begins by drawing the energy~$E$ of the incident particle from the input energy spectrum. For dark matter events, this is simply the recoil spectrum eq.~\eqref{eq:dRdE} (as in fig.~\ref{fig:spectrum}), while the background distributions are discussed in section~\ref{sec:backgrounds} (displayed in fig.~\ref{fig:backrates}). For this energy, we find the total number of quanta $\Nquanta$ by drawing from a Normal distribution with mean $\nquanta$ and variance $F\cdot \nquanta$, where $F=0.05$ is the Fano factor~\cite{Doke:1976zz}. We next separate $\Nquanta$ into excitons and ions. The number of ions $\Nion$ is drawn from a binomial distribution with $\Nquanta$ trials and a probability $(1+\nex/\nion)^{-1}$ that an ion is produced. The number of excitons $\Nex$ is simply $\Nex=\Nquanta-\Nion$. Our expressions for~$n_{\gamma}(E)$ and~$n_e(E)$ also include the effect of recombination fluctuations; we assume that the number of ions that recombine~$N^{\mathrm{recom}}_{\mathrm{i}}$ follows a Normal distribution with mean~$r \Nion$ and variance $\sigma_R^2=(1-r)C \Nion^2$, where $C=0.0056$~\cite{Lenardo:2014cva}. Our final result is that $n_{e}(E)=\Nion-N^{\mathrm{recom}}_{\mathrm{i}}$ and $n_{\gamma}(E)=f_l(\Nex+N^{\mathrm{recom}}_{\mathrm{i}})$, where $f_l$ is a quenching factor.  The Monte Carlo process is the same for both nuclear and electronic recoils; the difference is that $\nquanta$, $\nex$, $\nion$, $r$ and $f_l$ differ for nuclear and electronic recoils. The calculation of the mean quantities $\nquanta$, $\nex$, $\nion$, $r$ and the quenching factor $f_l$ are described in appendix~\ref{app:meanyields}. There is also a small difference between gamma- and beta-electronic interactions that we account for by rescaling $n_{\gamma}(E)$ and $n_{e}(E)$ calculated for a beta-interaction to obtain the result for the gamma-interaction. We perform this rescaling at this point, after the intrinsic fluctuations. Further details are given in appendix~\ref{app:meanyields}. 

We next include detector fluctuations in our calculation of~$\Sone$ and~$\Stwob$. For~$\Sone$, the number of photoelectrons~$N_{\mathrm{PE}}$ is drawn from a binomial distribution with~$n_{\gamma}(E)$ trials and success probability~$g_1$. The final result for~$\Sone$ also accounts for the PMT resolution:~$\Sone$ is drawn from a Normal distribution with mean~$N_{\mathrm{PE}}$ and variance~$\sigma^2_{\mathrm{PMT}} N_{\mathrm{PE}}$. For~$\Stwob$, the number of electrons~$N_e$ that are extracted from the liquid to the gas phase follows a binomial distribution with~$n_{e}(E)$ trials and success probability~$\epsilon$. To account for the amplification factor from converting ionisation electrons to photoelectrons, we draw from a Normal distribution with mean~$0.43\cdot Y\cdot N_e$ and variance~$\sigma^2_{\rm{PE_b}} N_e$ (the factor $0.43$ is the factor that relates $\Stwo$ and $\Stwob$). 

When generating the $\Sone$ signal from the inelastic scattering process, we combine the number of photons from the nuclear recoil with the photons from the de-excitation gamma-ray with energy $E^*$ after the intrinsic fluctuations (for which the two processes are treated independently) but before including the detector fluctuations. An analogous procedure is performed for $\Stwob$ except it is electrons that we combine before including the detector fluctuations.

\subsection{Two-phase xenon detector parameters}
\label{sec:detect}

The LUX and XENON collaborations have produced the most sensitive xenon detectors. The current LUX detector has a fiducial mass of $118~\mathrm{kg}$ and they have collected an exposure of 0.028 tonne-years~\cite{Akerib:2013tjd}, with an ultimate aim of collecting around 0.2 tonne-years~\cite{Akerib:2012ys}. The applied drift field of $181~\mathrm{V/cm}$ is lower than in previous experiments. For instance, ZEPLIN-III~\cite{Akimov:2011tj} and XENON100~\cite{Aprile:2012nq} had fields of $3400~\mathrm{V/cm}$ and $530~\mathrm{V/cm}$ respectively. However, the LUX light collection efficiency is much higher than in previous detectors, corresponding to a value of $g_1\approx0.12~\mathrm{PE}/\gamma$~\cite{Szydagis:2014xog,Huang2015}. Unfortunately the electron extraction efficiency is lower than was anticipated, with $\epsilon\approx 50\%$~\cite{Szydagis:2014xog,Huang2015}. The amplification factor is $Y\approx 24.6~\mathrm{PE}/e$~\cite{Akerib:2013tjd} so that $g_2\approx12~\mathrm{PE}/e$. The energy resolutions are $\sigma_{\mathrm{PMT}}\approx0.5~\mathrm{PE}/\gamma$ and $\sigma_{\mathrm{PE_b}}\approx 3.6~\mathrm{PE}/e$~\cite{Dobi:2014wza}.

XENON1T is the successor to XENON100. It will have a fiducial mass of approximately $1~\mathrm{tonne}$, a design drift field of~$1000~\mathrm{V/cm}$ and a light collection efficiency similar to LUX~\cite{Aprile:2012zx}. It is expected that the extraction efficiency $\epsilon$ will be 100\%, as achieved in XENON100. The amplification factor and resolutions will be similar to those in LUX and XENON100~\cite{Aprile:2013blg}. 

The follow-up to LUX is LZ~\cite{Malling:2011va,Akerib:2015cja}, with a projected fiducial mass of approximately $5.6~\mathrm{tonnes}$ and a drift field of $700~\mathrm{V/cm}$~\cite{Kudryavtsev:2015vja}. XENONnT is the successor of XENON1T and is designed to have similar characteristics as XENON1T but with a total mass of approximately $7~\mathrm{tonnes}$~\cite{Aprile:2014zvw}. As XENONnT and LZ will run for a number of years, an exposure of 15 tonne-years is readily achievable. Finally, there are plans for DARWIN, a much larger experiment with a fiducial mass of around 20 tonnes~\cite{Baudis:2012bc}. Studies assuming a drift field of 500~V/cm and an ultimate exposure of 200 tonne-years have been performed~\cite{Schumann:2015cpa}. This large exposure gives an indication of the ultimate reach of xenon detectors.

Future collaborations will obviously aim to optimise their respective detectors. It may be difficult to increase or even maintain the light collection efficiency because larger detectors collect a smaller fraction of the scintillation signal. LZ's proposal is that~$g_1>0.075$~\cite{Akerib:2015cja} and DARWIN studies have assumed the value reached in LUX~\cite{Schumann:2015cpa}. It should be possible to maintain an extraction efficiency close to unity and the amplification factor may be as large as $Y=50~\mathrm{PE}/e$~\cite{Akerib:2015cja}. 

In the remainder of this paper, we show results for two benchmark scenarios, {\it XenonA200} and {\it XenonB1000}, which should bracket the expected performance of upcoming experiments:
\begin{itemize}
\item {\it XenonA200} corresponds to a detector with lower drift field and lower extraction efficiency.  We assume a drift field of $200~\mathrm{V/cm}$ and the parameters $g_1=0.07~\mathrm{PE}/\gamma$, $\epsilon=50\%$, $Y=25~\mathrm{PE}/e$ so that $g_2=12.5~\mathrm{PE}/e$.
\item {\it XenonB1000} corresponds to a detector with a higher drift field and perfect extraction efficiency. We assume a drift field of $1000~\mathrm{V/cm}$ and the parameters $g_1=0.12~\mathrm{PE}/\gamma$, $\epsilon=100\%$, $Y=50~\mathrm{PE}/e$ so that $g_2=50~\mathrm{PE}/e$.
\end{itemize}
In both scenarios, we assume that $\sigma_{\mathrm{PMT}}=0.5~\mathrm{PE}/\gamma$ and $\sigma_{\mathrm{PE_b}}= 3.6~\mathrm{PE}/e$ and that $\Stwo_{\mathrm{b}}=0.43\times \Stwo$. Our benchmark exposure is 15 tonne-years unless stated otherwise. Finally, we assume that all measurement efficiencies are $100\%$ since the signals of interest are far from thresholds (where the efficiencies begin to deviate from $100\%$).

\subsection{Observable signals and their rate}

\begin{figure}[t!]
\centering
\includegraphics[width=0.49\columnwidth]{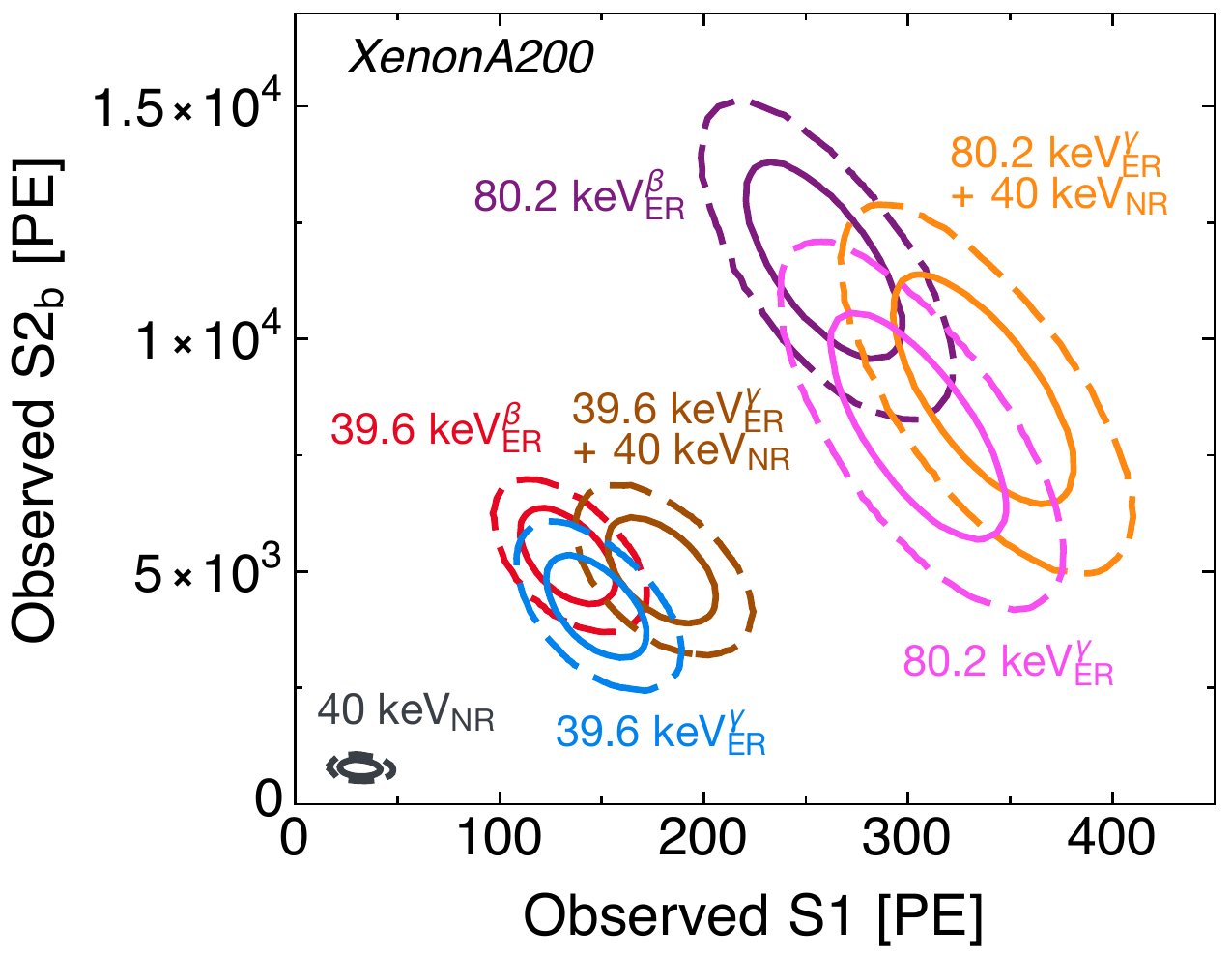} 
\includegraphics[width=0.49\columnwidth]{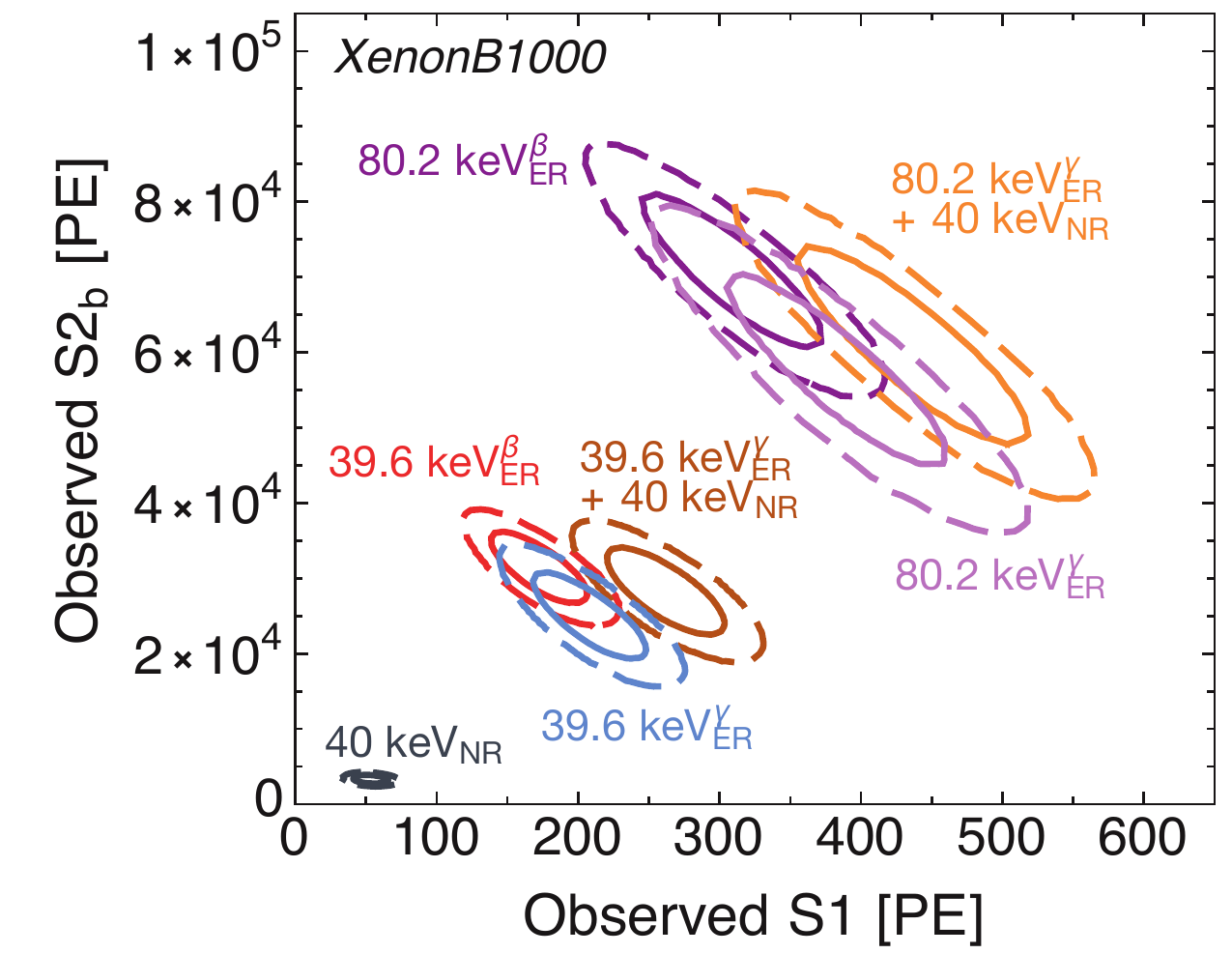} 
 \caption{The solid and dashed contours show where 68\% and 95\% of events occur in terms of the observable scintillation signals~$\Sone$ and~$\Stwob$ for the fixed input energies indicated. The left and right panels show results for the {\it XenonA200} and {\it XenonB1000} benchmark scenarios. The $\mathrm{keV}^{\beta}_{\mathrm{ER}}$, $\mathrm{keV}^{\gamma}_{\mathrm{ER}}$ and $\mathrm{keV}_{\mathrm{NR}}$ labels indicate that the energy originates from a beta-electronic, gamma-electronic and nuclear recoil. The brown and orange contours show the signal region for an event that occurs for inelastic scattering, which includes energy from both a nuclear recoil and a de-excitation photon. Inelastic signal events have higher $\Sone$ and $\Stwo$ values than for elastic scattering events, for which the search window is typically $\Sone\leq30~\mathrm{PE}$. All of the contours are tilted because of recombination fluctuations. Note the change of scale on both axes between the two panels.
 \label{fig:S1S2plot}}
 \end{figure}
 
Having described our procedure for generating light and charge signals, we can proceed to generate the observable signals for our two benchmark scenarios. The solid and dashed contours in figure~\ref{fig:S1S2plot} show where 68\% and 95\% of events occur in the $\Sone$\,-\,$\Stwob$ plane for fixed input energies. The left and right panels correspond to the {\it XenonA200} and {\it XenonB1000} benchmark scenarios, respectively. The $\mathrm{keV}^{\beta}_{\mathrm{ER}}$, $\mathrm{keV}^{\gamma}_{\mathrm{ER}}$ and $\mathrm{keV}_{\mathrm{NR}}$ labels in figure~\ref{fig:S1S2plot} indicate that the energy originates from a beta-electronic, gamma-electronic and nuclear recoil, respectively. The black contours show the signal region for a nuclear recoil of~$40~\mathrm{keV}$, the red and blue contours show the signal region for a $39.6~\mathrm{keV}$ electronic event induced by a beta- and gamma-ray respectively, and the purple and pink contours show the signal region for a $80.2~\mathrm{keV}$ electronic event induced by a beta- and gamma-ray, respectively. The brown and orange contours show the signal region for an event that occurs for inelastic scattering: in this case the nuclear recoil energy is $40~\mathrm{keV}$ and the gamma-electronic energy is $39.6~\mathrm{keV}$ and $80.2 ~\mathrm{keV}$, corresponding to the energies of the photon emitted in the de-excitation of the $\xea$ and $\xeb$ isotopes, respectively.

We first discuss the features common to both panels. These features are well known properties of two-phase xenon detectors and are reviewed in much more detail in~\cite{Chepel:2012sj}. We see that a nuclear recoil typically produces a much smaller $\Sone$ and $\Stwob$ signal compared to an electronic recoil of the same energy. The usual XENON100 and LUX dark matter searches for elastic scattering define a $\Sone$ search window up to $30~\mathrm{PE}$~\cite{Aprile:2012nq,Akerib:2013tjd}; we see that for inelastic signals, we will have to consider much higher values of $\Sone$. The difference between a gamma- and beta-interaction of the same energy is relatively small, $\mathcal{O}(10\%)$, for both~$\Sone$ and~$\Stwob$. It is also apparent that adding a nuclear recoil to an electronic recoil only slightly increases the $\Sone$ and $\Stwob$ signals compared to a pure electronic recoil. Both panels show that the contours are tilted, matching the behaviour observed with real data (see e.g.~\cite{Xiao:2015psa}). This is especially obvious in the events where the $\Sone$ and $\Stwob$ signals are dominated by electronic recoils while for nuclear recoils, the tilt is much smaller. The origin of the tilt is well-known: it is a result of recombination fluctuations, which are 100\% anti-correlated in scintillation~$(\Sone)$ and ionisation~$(\Stwo)$~\cite{Aprile:2007qd}. In contrast, the detector fluctuations smear along constant $\Sone$ and constant $\Stwo$ only. The tilt is smaller for the nuclear recoil region because the detector fluctuations are larger than recombination fluctuations.
 
We next discuss the features that differ between the panels. The first important difference is that the~$\Sone$ and~$\Stwob$ signals are much larger in the right panel for all configurations (note that the scales differs in the two panels). The larger~$\Stwob$ signal is a result of two effects. The first is that the extraction efficiency~$\epsilon$ and amplification factor~$Y$ are both twice and therefore~$g_2$ is four times larger for {\it XenonB1000}. If this were the only effect, then $\Stwob$ would be four times larger for the same input energy. In fact, we see that~$\Stwob$ is around six times larger in the right panel. This is because the larger drift field also increases~$\Stwob$. The larger drift field reduces the recombination fraction so more of the ions survive to form the~$\Stwob$ signal. The higher drift field also reduces the~$\Sone$ signal for the same reason; fewer ions recombine producing fewer prompt scintillation light. However,~$g_1$ is~70\% larger for {\it XenonB1000}, which compensates for the reduction from the higher drift field. This is why the~$\Sone$ values are actually~$\sim20\%$ larger in the right panel. Finally, an important difference is that there is less overlap between the contours from an inelastic signal (brown and orange contours) and the contours from a potential beta-background source (red and purple contours) for {\it XenonB1000}. This greater separation is again an effect of the higher drift field. We will see in section~\ref{sec:results} that this better discrimination is ultimately responsible for the greater sensitivity of the {\it XenonB1000} benchmark scenario.

 \begin{figure}[t!]
\centering
\includegraphics[width=0.49\columnwidth]{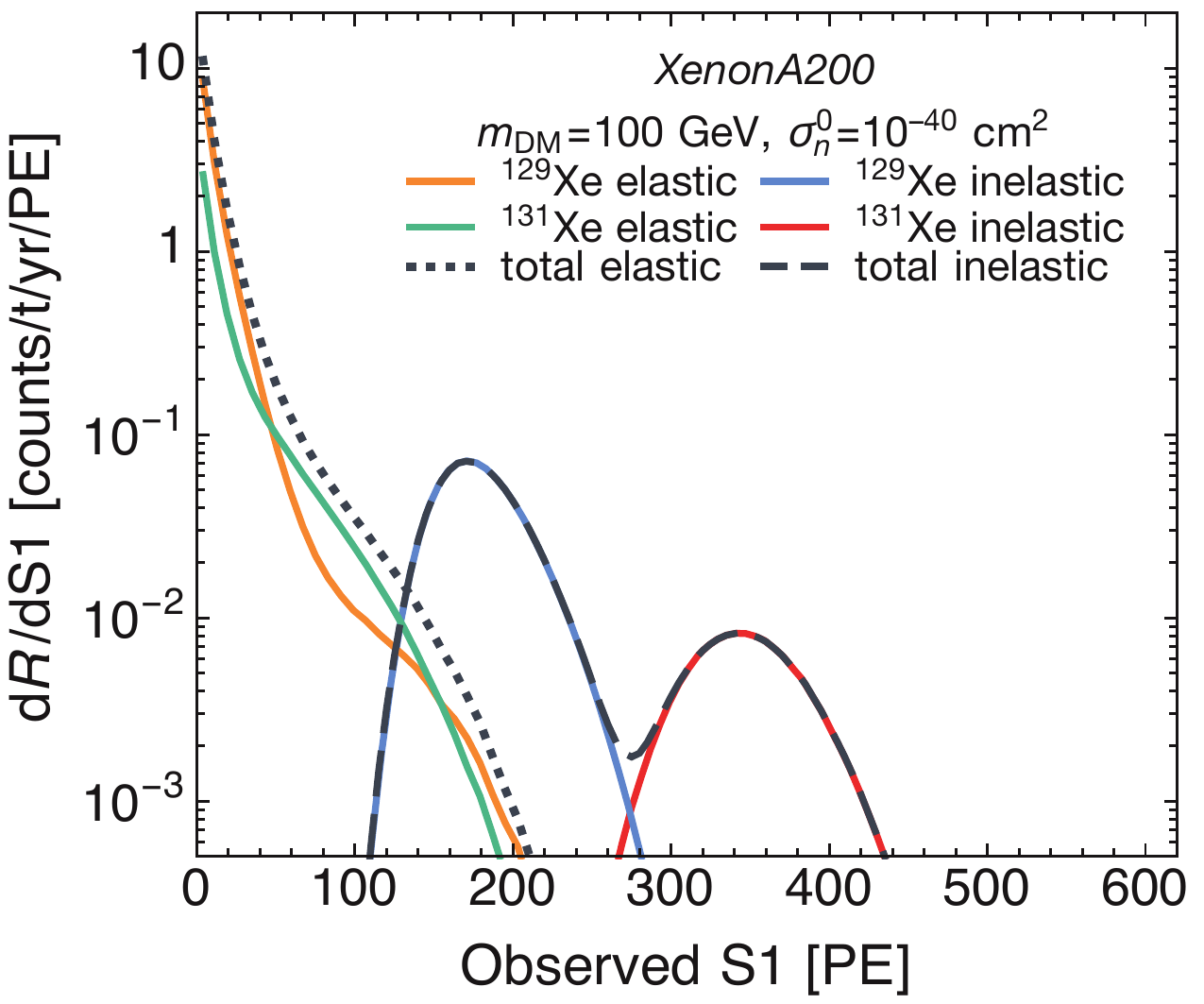} 
\includegraphics[width=0.49\columnwidth]{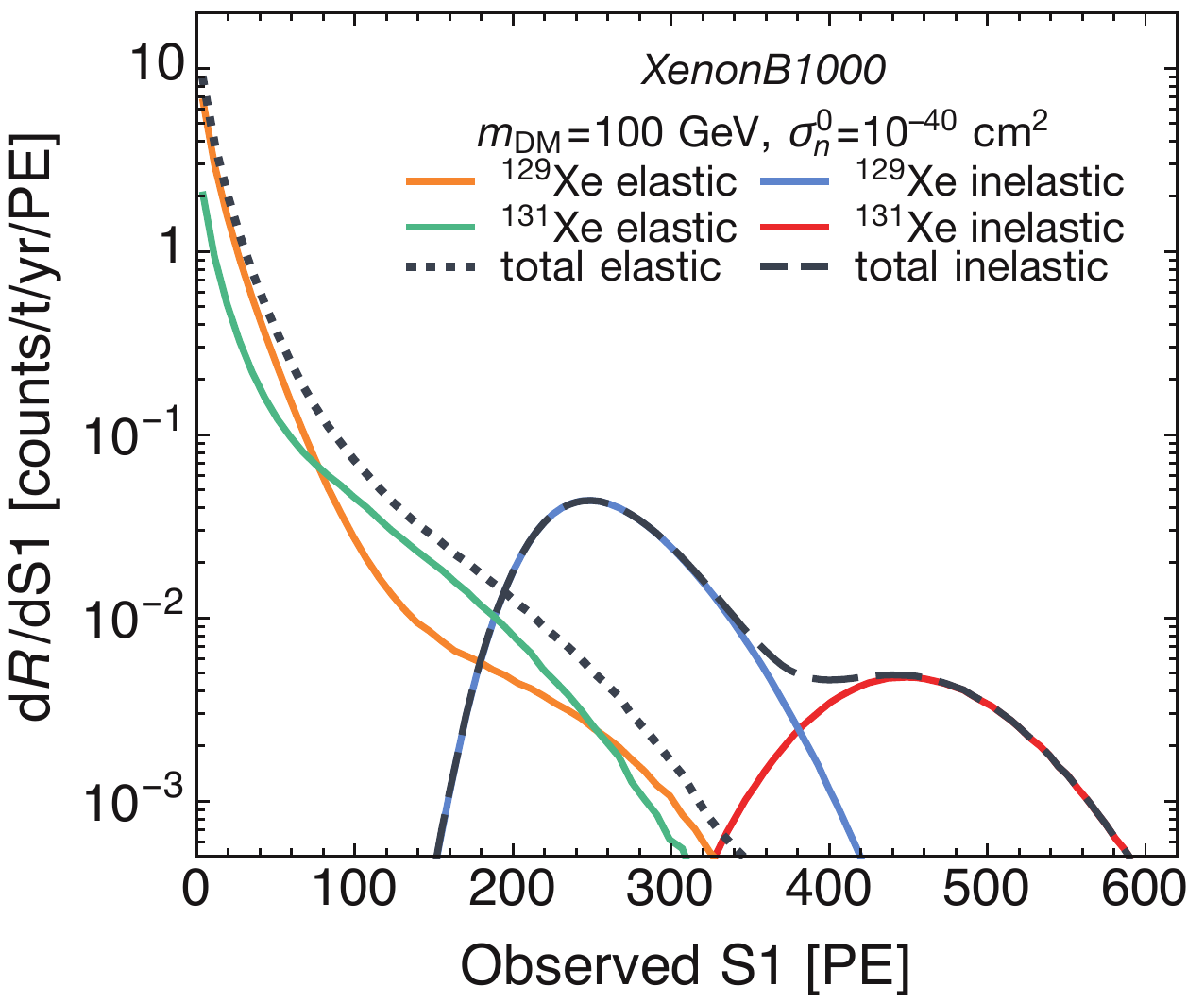} 
 \caption{The left and right panels show the recoil spectra for the elastic and inelastic processes in terms of~$\Sone$ for the two benchmark scenarios {\it XenonA200} and {\it XenonB1000} described in section~\ref{sec:detect}. These spectra are for~$\mDM=100~\mathrm{GeV}$ and~$\sigma_n^0=10^{-40}~\mathrm{cm}^2$. While the elastic spectrum falls off rapidly, the inelastic spectrum has two distinct peaks. The first peak is dominated by the $39.6~\mathrm{keV}$ de-excitation from~$\xea$ while the second peak is dominated by the $80.2~\mathrm{keV}$ de-excitation from~$\xeb$. Similar to the left panel of figure~\ref{fig:spectrum}, the inelastic rate is suppressed by about two (three) orders of magnitude with respect to the elastic rate for the~$\xea$ $\left(\xeb\right)$ isotope. \label{fig:dRdS1spect}}
 \end{figure}

We previously only gave the differential event rate in terms of the xenon recoil energy $\Er$ (cf.~eq.~\eqref{eq:dRdE} and figure~\ref{fig:spectrum}). We are now in a position to calculate the event rate in terms of the observable quantities $\Sone$ and $\Stwob$. The differential event rate is~\cite{Aprile:2012vw}
\begin{equation}
\label{eq:dRdS1}
\frac{d^2 R}{d \Sone\, d \Stwob}=\int d \Er \frac{dR}{d \Er} \mathrm{pdf}(\Sone, \Stwob|\Er)\;,
\end{equation}
where $dR/d\Er$ is eq.~\eqref{eq:dRdE} and $\mathrm{pdf}(\Sone, \Stwob|\Er)$ is the probability density function, which we generate with the Monte Carlo process described in section~\ref{sec:gensignal}.

As an example of our results, we show in figure~\ref{fig:dRdS1spect} the differential event rate in terms of~$\Sone$ for $\mDM=100~\mathrm{GeV}$ and~$\sigma_n^0=10^{-40}~\mathrm{cm}^2$ (additionally, the results for $\mDM=1000~\mathrm{GeV}$ and~$\sigma_n^0=10^{-39}~\mathrm{cm}^2$ are shown in figure~\ref{fig:drdsrates}). The left and right panels correspond to the ${\it XenonA200}$ and ${\it XenonB1000}$ benchmark scenarios, respectively. The $dR/d\Sone$ spectrum is obtained by additionally integrating eq.~\eqref{eq:dRdS1} over $\Stwob$. The colour scheme of the lines matches figure~\ref{fig:spectrum}: the total elastic and inelastic spectrum are shown by the black dotted and black dashed lines respectively. The orange and green lines show the contribution of $\xea$ and $\xeb$ to the elastic spectrum, while the blue and red lines show the contribution of $\xea$ and $\xeb$ to the inelastic spectrum. As in the left panel of figure~\ref{fig:spectrum}, the inelastic rate for scattering with~$\xea$ $\left(\xeb\right)$ is suppressed by about two (three) orders of magnitude with respect to the elastic rate.

Figure~\ref{fig:dRdS1spect} shows the well known fact that the elastic spectrum falls off rapidly with~$\Sone$. In contrast, the inelastic spectrum has two distinct peaks whose origin is clear. The first peak is due to the $39.6~\mathrm{keV}$ de-excitation photon from the~$\xea$ isotope while the second peak is from the $80.2~\mathrm{keV}$ de-excitation photon from the~$\xeb$ isotope. The peak $\Sone$ values agree with the values shown in figure~\ref{fig:S1S2plot}. The peak differential rate is slightly higher for the inelastic process in the left panel (corresponding to {\it XenonA200}) because the spectrum is slightly more peaked in $\Sone$ (the integrated rate is the same).

\section{Background rates}
\label{sec:backgrounds}

Our ultimate aim is to assess the discovery potential of the inelastic signal. In order to do this, the background signals must be quantified. A comprehensive study of the backgrounds for tonne-scale xenon detectors was performed in~\cite{Baudis:2013qla} and similar rates and distributions were also presented by the LZ~collaboration~\cite{Beltrame:2014,Akerib:2015cja}. We summarise the relevant results for our study and refer the reader to~\cite{Baudis:2013qla,Beltrame:2014,Akerib:2015cja} for further details.

\begin{figure}[t!]
\centering
\includegraphics[width=0.49\columnwidth]{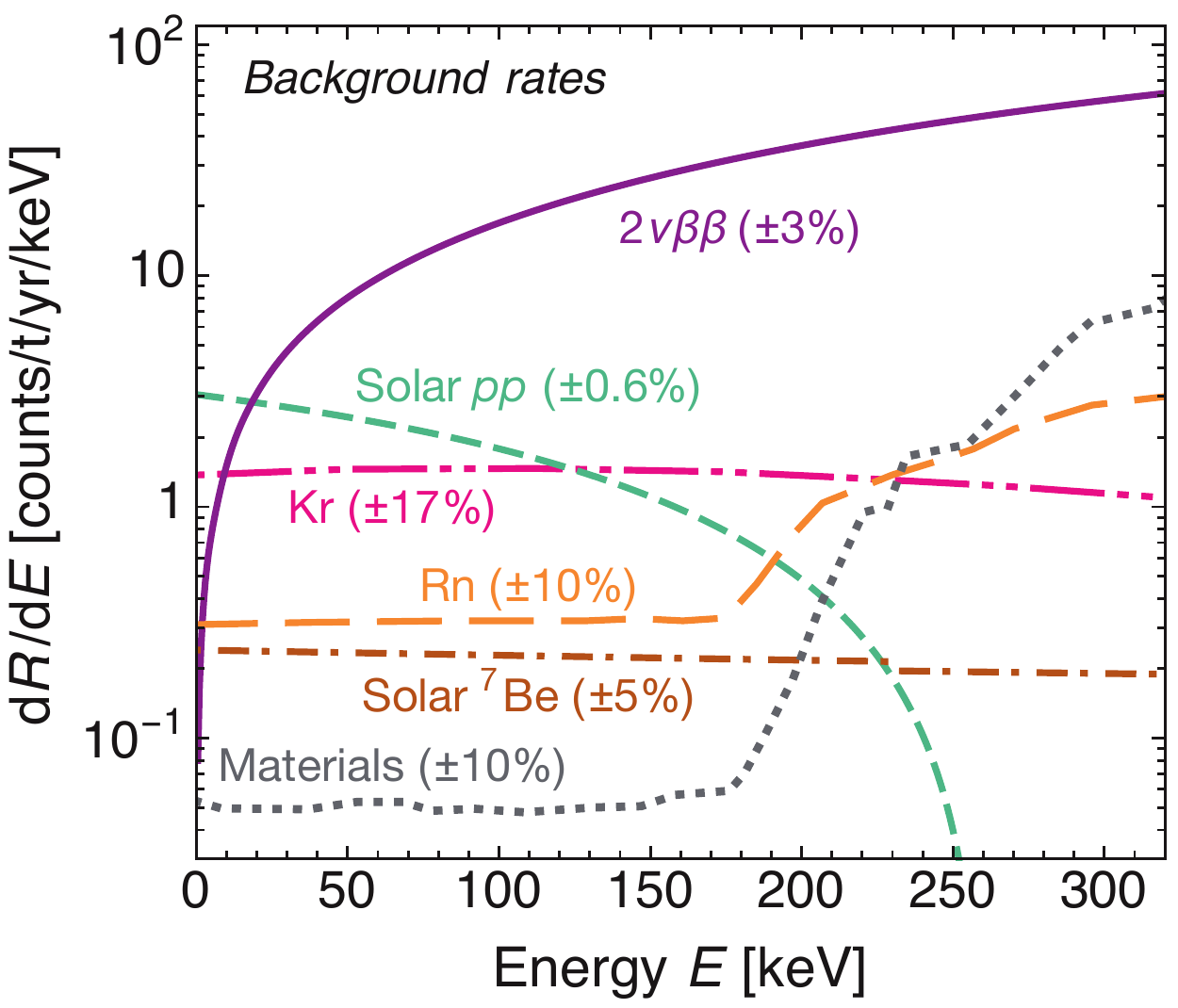} 
 \caption{The main backgrounds and their rates for tonne-scale xenon experiments. The dominant background rate is from the $2\nu\beta\beta$-decay of the $\xec$ isotope, which has an abundance of $8.86\%$ in natural xenon. There are two irreducible backgrounds from~$pp$ and~$^7\mathrm{Be}$ solar neutrinos scattering on atomic electrons. The percentage figure associated with each background is the uncertainty in the normalisation of that rate.
 \label{fig:backrates}}
 \end{figure}

The dominant backgrounds are those that produce electronic recoils in the signal range of interest, $\Sone\leq600~\mathrm{PE}$, which corresponds to an energy range of approximately $0-300$~keV.  As discussed in~\cite{Baudis:2013qla,Beltrame:2014}, the background rates that dominate in order of decreasing importance are the $2\nu\beta\beta$-decay of $\xec$, elastic neutrino-electron scattering from~$pp$ and~$^7\mathrm{Be}$ solar neutrinos, decays of~$\kr$ and~$\rn$ and finally, radioactivity from detector materials. All of these backgrounds are beta-electronic sources~\cite{Malling:2014wza}. The background rates and their uncertainties used in this study are shown in figure~\ref{fig:backrates}. We comment on each of the rates in turn.

We recalculated the background rates from the $2\nu\beta\beta$-decay of~$\xec$ and the elastic scattering from~$pp$ and~$^7\mathrm{Be}$ solar neutrinos using updated parameters. For the $2\nu\beta\beta$-decay, the neutrinos escape the detector while the beta-particles contribute to the background rate. We calculated the rate assuming the $\xec$ abundance is that of natural xenon: $8.86\%$. We use the most accurate measurement of the $\xec$ half-life by EXO-200, who found~$T_{1/2}=(2.165\pm0.059)\times10^{21}~\mathrm{yr}$~\cite{Albert:2013gpz}, where we only quote the dominant systematic error. We use the distribution of the summed energies of the beta-particles from~\cite{Kotila:2012zza}. Our~$2\nu\beta\beta$ rate is in good agreement with that shown in~\cite{Baudis:2013qla,Beltrame:2014}. The dominant uncertainty of this rate is from~$T_{1/2}$, at the level of~$3\%$. In comparison, the abundance of the~$\xec$ isotope can be measured with a~$0.05\%$ accuracy~\cite{brownsimgen}.

The largest flux of solar neutrinos is from the $pp$ chain. The~$pp$ flux measured by Borexino, $(6.6\pm0.7)\times10^{10}~\mathrm{cm}^{-2}\mathrm{s}^{-1}$~\cite{Bellini:2014uqa}, is in good agreement with the prediction of the Standard Solar Model (SSM) $6.03\times(1\pm0.06)\times10^{10}~\mathrm{cm}^{-2}\mathrm{s}^{-1}$~\cite{Serenelli:2011py}. The SSM prediction is well understood so we use the theoretical flux and error in our calculation. The second largest rate from solar neutrinos is from~$^7\mathrm{Be}$ neutrinos. Borexino measured a flux of $(4.84\pm0.24)\times10^{9}~\mathrm{cm}^{-2}\mathrm{s}^{-1}$~\cite{Bellini:2011rx}, also in good agreement with the SSM prediction~\cite{Serenelli:2011py}. We use the measured value and error in our calculation. We use the neutrino-electron scattering cross-section from~\cite{Marciano:2003eq}. Finally, we use the electron neutrino survival probabilities listed in~\cite{Bellini:2013lnn}. Our~$pp$ spectrum agrees well with~\cite{Baudis:2013qla,Beltrame:2014}. Our~$^7\mathrm{Be}$ spectrum agrees well with~\cite{Beltrame:2014} but is about a factor~1.6 smaller than in~\cite{Baudis:2013qla}. The origin of the discrepancy is unclear.\footnote{There is a typo in the neutrino-electron cross-section formula in~\cite{Baudis:2013qla} (the last term has the wrong dimensions) but this is not the origin of the discrepancy.} The third largest rate from solar neutrinos is from $^{13}\mathrm{N}$ neutrinos. This rate is approximately 300 times smaller than the $pp$ rate so it is a good approximation to ignore the contribution to the rate from all solar neutrinos except the~$pp$ and~$^7\mathrm{Be}$ neutrinos.

We use the~$\kr$,~$\rn$ and detector material background rates from the study in~\cite{Baudis:2013qla}, which assumes a~$\kr$ contamination of~$0.1~\mathrm{ppt}$ and a~$\rn$ level of~$0.1~\mu\mathrm{Bq}/\mathrm{kg}$. While the~$\rn$ rate is similar in the LZ study, the~$\kr$ rate is a factor four smaller~\cite{Massoli:2015}. We use the result from~\cite{Baudis:2013qla} because the assumptions entering the calculation are clearer. XENON100 and EXO-200 have measured the~$\kr$ contamination and~$\rn$ level with an accuracy of~$17\%$~\cite{Lindemann:2013kna} and~$10\%$~\cite{Albert:2013gpz} respectively, and we assume the same accuracy will be achieved in the future.

The detector material background rate is reduced by self-shielding of the liquid xenon so larger detectors, which have more xenon with which to shield, have a smaller rate. The rate reported here was for a DARWIN study and assumes a 14~tonne fiducial mass. The rate for LZ with a 5.6~tonne fiducial mass is about three times larger~\cite{Beltrame:2014}. As before, we use the result from~\cite{Baudis:2013qla} because the assumptions entering the calculation are clearer. In any case, as this rate is always subdominant, a factor three difference has an almost negligible impact on our results. Both the material and $\rn$ background rates begin to increase after $\sim170~\mathrm{keV}$ however they always remain subdominant to the~$2\nu\beta\beta$-decay rate~\cite{Baudis:2013qla}. The detector material rate for XENON1T is predicted to~$10\%$~\cite{Massoli:2015} and we assume the same accuracy will be achieved in future experiments.

Before leaving this sub-section, we briefly comment on background sources that do not produce electronic recoils. It is also possible that neutrons may directly excite the~$\xea$ and~$\xeb$ isotopes creating another background source (that neutrons can excite the signal is actually an advantage since at least in principle, it allows the signal region to be calibrated in a real detector). The self-shielding of the liquid xenon and a dedicated muon-veto system to reject muon-induced neutrons means that the neutron rate can be reduced to less than $5\times10^{-5}~\mathrm{counts/t/yr/keV}$ for single-scatter neutrons that elastically scatter of xenon~\cite{Schumann:2015cpa}. For comparison, the dark matter signal rate that we consider in this paper is $\sim10^{-2}~\mathrm{counts/t/yr/keV}$ for $\sigma_n^0\simeq10^{-40}~\mathrm{cm}^2$ (cf.~fig.~\ref{fig:drdsrates}). Although there are no detailed studies that discuss inelastic neutron scattering (and such a study is beyond the scope of this paper), the inelastic neutron scattering cross-section is generally of the same order of magnitude as the elastic scattering cross-section~\cite{NDS}. Therefore, we assume that the inelastic neutron scattering rate is similar to the elastic scattering rate and is therefore always significantly smaller than the dark matter signal rate, so we ignore this background contribution in our study. A more detailed study to confirm this assumption is desirable.

\subsection{Comparing background and signal rates}
\label{subsec:comp}
 
\begin{figure}[t!]
\centering
\includegraphics[width=0.49\columnwidth]{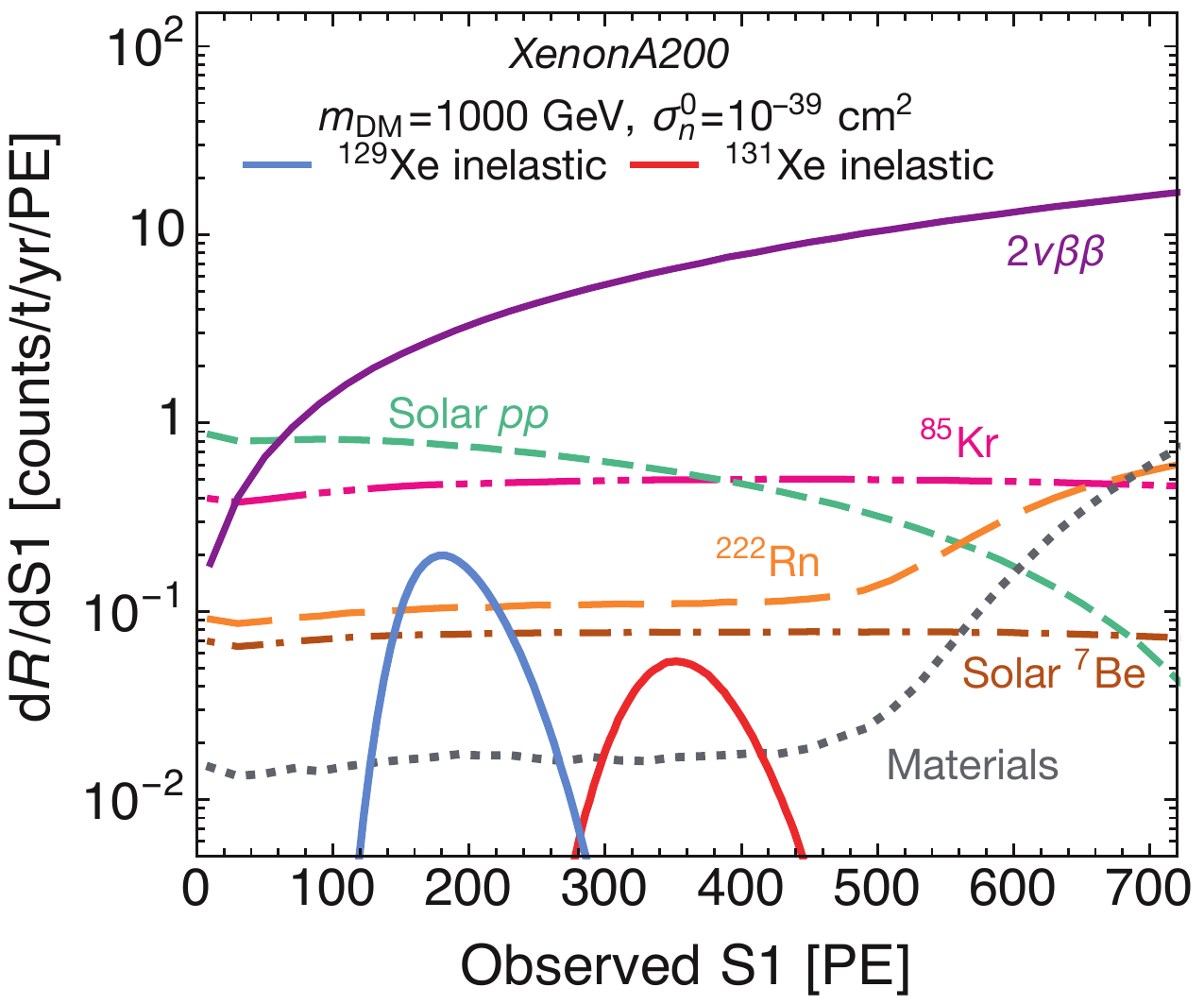} 
\includegraphics[width=0.49\columnwidth]{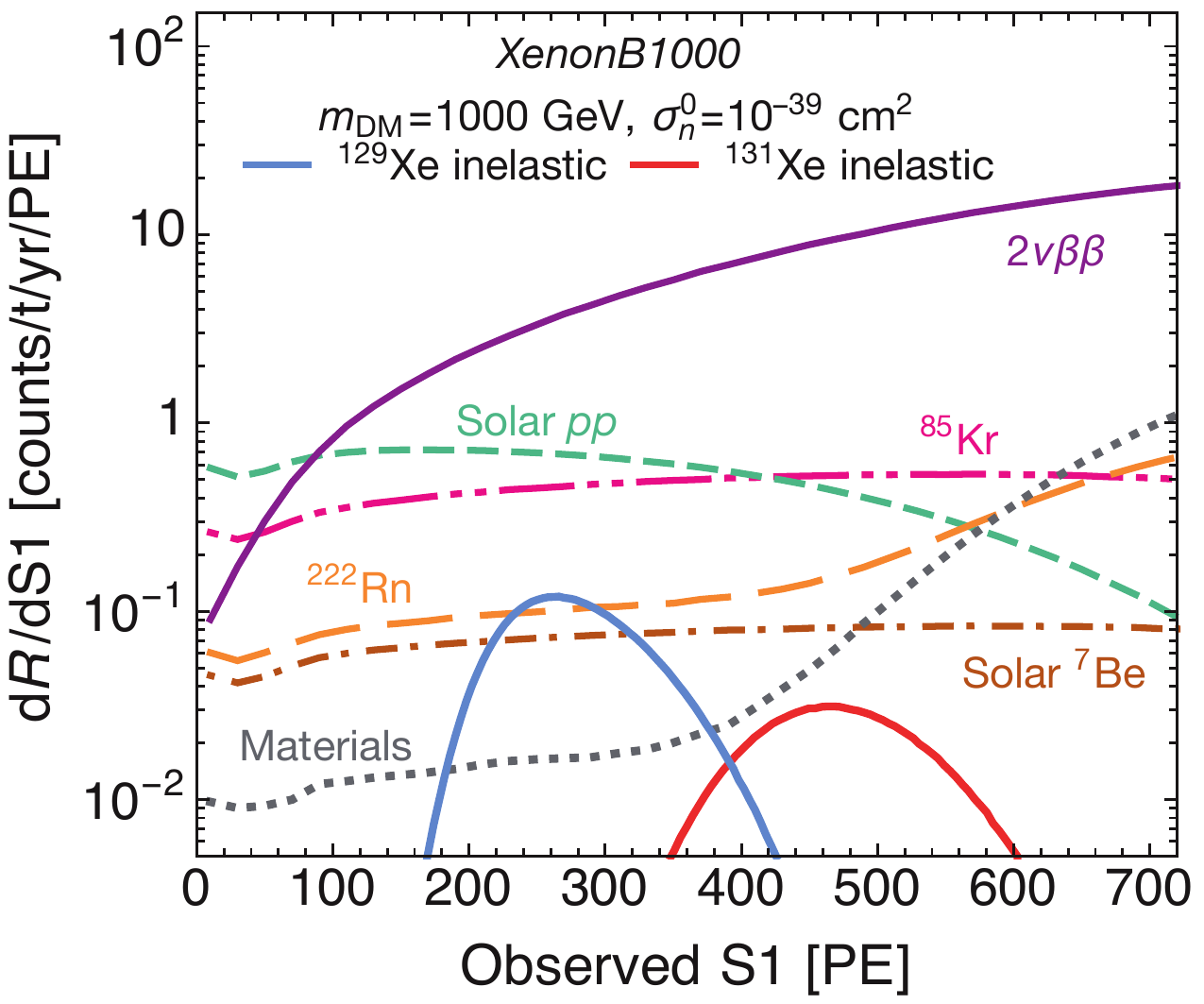}
 \caption{The left and right panels show the background and inelastic signal rates for the two benchmark scenarios {\it XenonA200} and {\it XenonB1000} described in section~\ref{sec:detect}. The~$\xea$ (blue line) and~$\xeb$ (red line) inelastic spectra are for~$\mDM=1000~\mathrm{GeV}$ and~$\sigma_n^0=10^{-39}~\mathrm{cm}^2$. The dominant background rate is from the $2\nu\beta\beta$-decay of the $\xec$ isotope (solid purple line), which is always at least~30 times larger than the dark matter signal. These panels demonstrate that observing the inelastic signal with a single-phase xenon experiment that only measures the~S1 signal will be very challenging because of the large background rate.
\label{fig:drdsrates}}
 \end{figure}

We now have everything to model the signal and the background for the {\it XenonA200} and {\it XenonB1000} detector scenarios. The left and right panels in figure~\ref{fig:drdsrates} show the background and signal rates as a function of~S1 for the two benchmark scenarios. The red and blue lines show the signal rate corresponding to inelastic scattering off the~$\xea$ and~$\xeb$ isotopes for~$\mDM=1000~\mathrm{GeV}$ and~$\sigma_n^0=10^{-39}~\mathrm{cm}^2$. Comparing the background rates in the left and right panels, we see only minor differences. In both cases the dominant background is from the $2\nu\beta\beta$-decay of~$\xec$ (solid purple line). The most obvious difference is in the rate from detector materials (dotted grey line), where for {\it XenonB1000}, the rate is higher for~$\Sone$ values corresponding to the $80.2~\mathrm{keV}$ de-excitation. The main point to take away from both panels of figure~\ref{fig:drdsrates} is that the signal rate is always at least 30 times smaller than the background rate. This demonstrates that observing this signal with a single-phase xenon experiment that only measures the~S1 signal will be very challenging. 

\begin{figure}[t!]
\centering
\includegraphics[width=0.49\columnwidth]{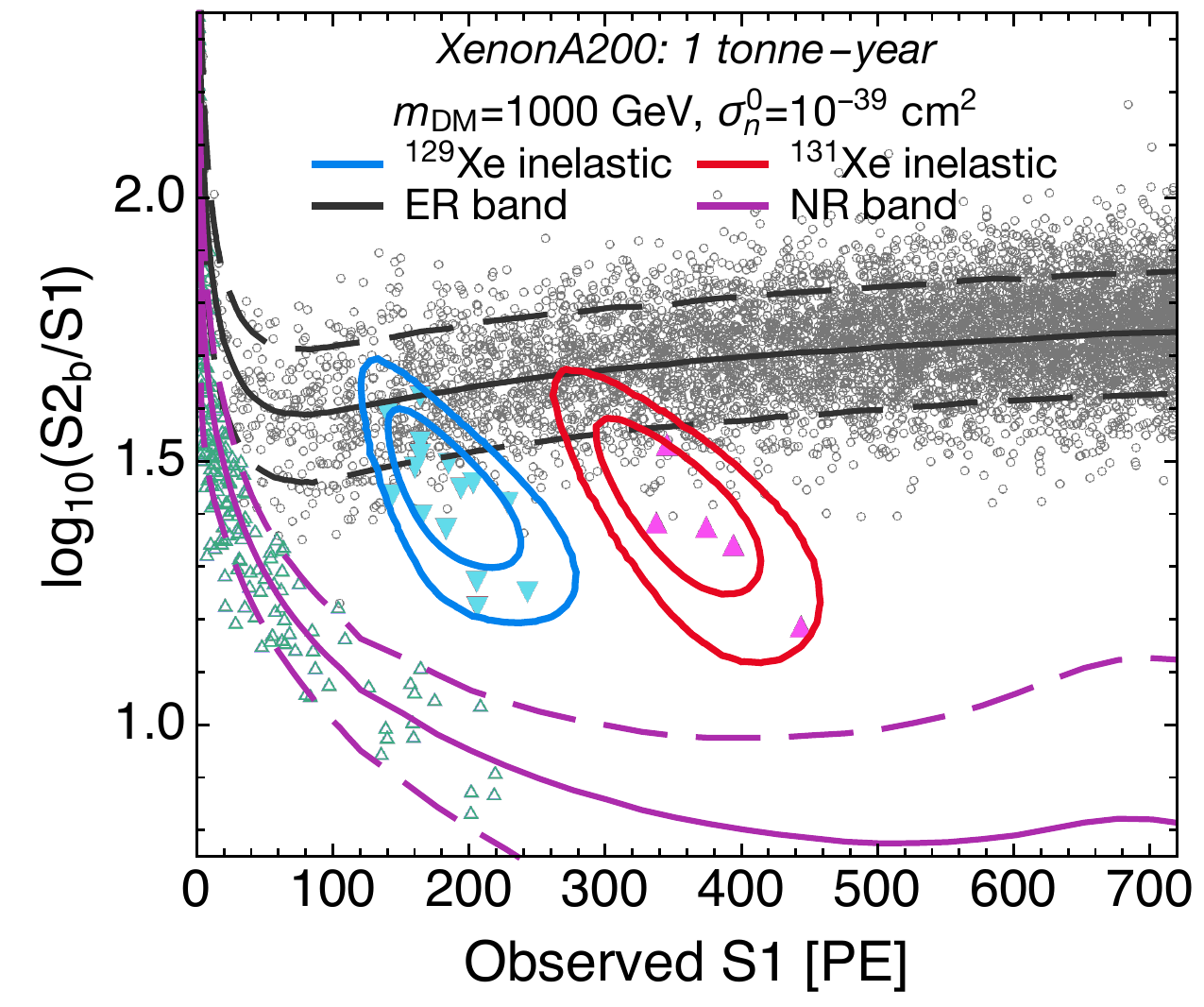} 
\includegraphics[width=0.49\columnwidth]{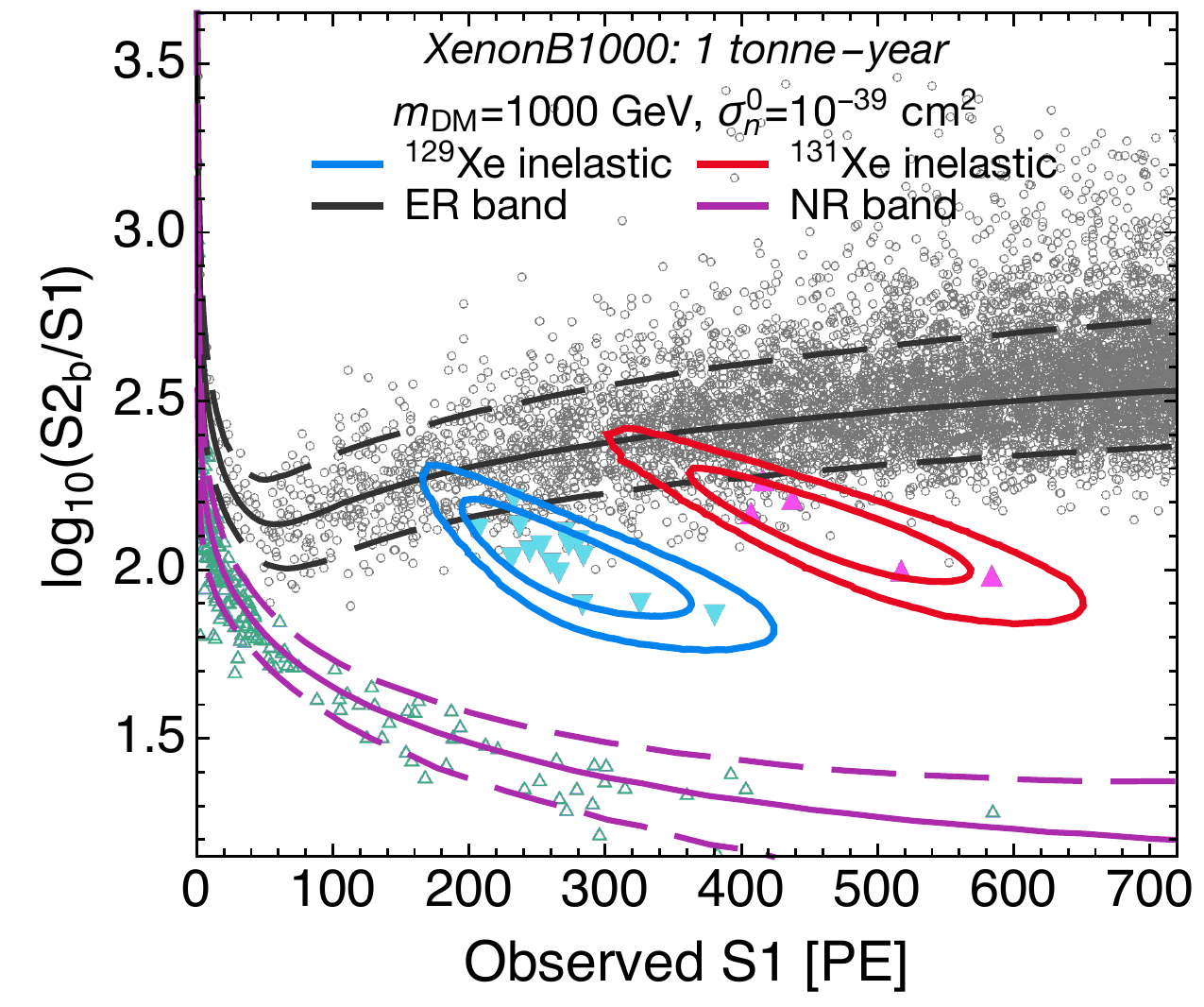}
 \caption{The left and right panels show a simulation of the background and signal regions for the {\it XenonA200} and {\it XenonB1000} benchmark scenarios described in section~\ref{sec:detect}. The black and purple bands show the electronic and nuclear recoil bands which contain~$80\%$ of the background and elastic scattering dark matter signals. The red and blue contours show where~68\% and~95\% of events occur for inelastic scattering with the~$\xea$ and~$\xeb$ isotopes for~$\mDM=1000~\mathrm{GeV}$ and~$\sigma_n^0=10^{-39}~\mathrm{cm}^2$. The circles and triangles show the simulated events expected for an exposure of one tonne-year. The open grey circles show the background events, the filled blue and filled red triangles show the inelastic events arising from inelastic scattering off the~$\xea$ and~$\xeb$ isotopes, while the open green triangles show the events from elastic scattering off xenon. Two-phase xenon experiments allow for some discrimination between the inelastic signal and the background events because the signal region extends below the electronic recoil band.
  \label{fig:StwoSone}}
 \end{figure}

Two-phase experiments provide additional information in the form of the~$\Stwo$ signal. In figure~\ref{fig:StwoSone} we therefore plot the signal and background distributions in the~$\log_{10}\left(\Stwob/\Sone \right)$ -- $\Sone$ plane traditionally used by two-phase xenon experiments. The black and purple lines show the electronic and nuclear recoil bands, respectively. The solid lines show the median while the dashed lines show $\pm1.28 \sigma$ around the median, such that~$10\%$ of events are  above and~$10\%$ below the dashed lines. The bands are calculated by passing a constant energy spectrum through our detector simulations for nuclear and beta-electronic recoils. The overall shape of the bands, and in particular that they separate at large~$\Sone$, matches the behaviour observed with real detectors (see e.g.~\cite{Dahl:2009nta}). The blue and red contours indicate where 68\% and 95\% of events occur for inelastic scattering off the~$\xea$ and~$\xeb$ isotopes for~$\mDM=1000~\mathrm{GeV}$ and~$\sigma_n^0=10^{-39}~\mathrm{cm}^2$, respectively. Unlike the contour regions shown in figure~\ref{fig:S1S2plot}, these contours are not elliptical but have a more extended shape. This shape change arises because figure~\ref{fig:StwoSone} includes the effect of all possible recoil energies of the nucleus while figure~\ref{fig:S1S2plot} was for a single nuclear recoil energy. Finally, the circles and triangles show the simulated events expected for an exposure of one tonne-year and the dark matter parameters mentioned above. The open grey circles show the electronic background events, which as expected from figure~\ref{fig:drdsrates}, become more abundant at higher values of~$\Sone$. The filled blue and filled red triangles show the inelastic events from the $39.6~\mathrm{keV}$ and $80.2~\mathrm{keV}$ de-excitation after scattering off the~$\xea$ and~$\xeb$ isotopes. The open green triangles show the events from elastic scattering off xenon, which are more abundant at smaller values of~$\Sone$.

Both panels of figure~\ref{fig:StwoSone} show that the signal and background distributions are slightly displaced. The displacement occurs for two reasons. The first is that nuclear recoils also have a lower~$\Stwob$ for the same~$\Sone$ compared to an electronic event (this is why the nuclear band is below the electronic recoil band) and the second is that gamma-electronic interactions have a higher~$\Sone$ and lower~$\Stwob$ than a beta-interaction of the same energy (as shown in figure~\ref{fig:S1S2plot}). Both effects mean that the inelastic signal region lies below the electronic recoil band. This displacement is crucial as it allows for some discrimination between signal and background events. This means that two-phase xenon detectors should have a significantly better sensitivity than single-phase detectors.

Finally, we discuss the differences between the two benchmark scenarios. All of the signals have larger~$\Sone$ and~$\Stwob$ values for the {\it XenonB1000} scenario because of the larger~$g_1$ and~$g_2$ values. The extent to which the signal regions extend below the electronic recoil band depends on the detector parameters, particularly the applied electric drift field. For {\it XenonA200} (left panel) which has a drift field of $200~\mathrm{V/cm}$, $78\%$ of the~$\xea$ inelastic events signal fall below the lower dashed line of the electronic recoil band. In comparison, for {\it XenonB1000} (right panel) where the drift field is $1000~\mathrm{V/cm}$, $92\%$ of the~$\xea$ inelastic events signal fall below this line. For the~$\xeb$ signal region, $92\%$ of the events fall below the lower dashed line of the electronic recoil band for both the {\it XenonA200} and {\it XenonB1000} benchmark scenarios so we expect a similar sensitivity to the inelastic signal from the~$\xeb$ isotope for these scenarios.

To demonstrate that the drift field is primarily responsible for the separation of the signal region and electronic recoil band, we repeated this analysis with the detector parameters of {\it XenonA200} but with a drift field of $1000~\mathrm{V/cm}$ instead of $200~\mathrm{V/cm}$. In this case, we found that $91\%$ of the~$\xea$ inelastic events signal fall below the lower dashed line of the electronic recoil band, similar to the $92\%$ obtained for {\it XenonB1000}. Similarly, for the detector parameters of {\it XenonB1000} but with a drift field of $200~\mathrm{V/cm}$ instead of $1000~\mathrm{V/cm}$, we found a value of $80\%$, similar to the value $78\%$ obtained for {\it XenonA200}. 

Figure~\ref{fig:StwoSone} was generated for~$\mDM=1000~\mathrm{GeV}$ and~$\sigma_n^0=10^{-39}~\mathrm{cm}^2$ but similar signal regions hold for other masses. This should not be too surprising since most of the~$\Sone$ and~$\Stwob$ signal originates from the de-excitation photon whose energy is always the same. The primary change is that the ratio of the~$\xea$ to~$\xeb$ rate is larger for smaller mass values (cf.~figures~\ref{fig:dRdS1spect} and~\ref{fig:drdsrates}). 

\section{Characterising the detection sensitivity}
\label{sec:discovery}

In this section we describe our method for characterising the sensitivity of two-phase xenon experiments to the inelastic scattering process. We will do this by calculating the `discovery limit' or as we will call it, the discovery reach. This was introduced in~\cite{Billard:2011zj} and has been used extensively to characterise the limiting effect of the neutrino background (see e.g.~\cite{Billard:2013qya}). We first describe the formalism behind this frequentist approach and then provide specific details of our calculation.

The discovery reach is the smallest cross-section for which~90\% of experiments make at least a~$3\sigma$ discovery of the signal under consideration. To calculate it, we make use of the frequentist test statistic for the discovery of a positive signal~\cite{Cowan:2010js}:
\begin{equation}
q_0=\begin{cases}
-2\ln \lambda(0) &\hat{\sigma}_n^0\geq 0\\
0 &\hat{\sigma}_n^0<0
\end{cases}
\end{equation}
where the profile likelihood ratio is
\begin{equation}
\lambda(0)=\frac{L(\sigma_n^0=0,\hat{\hat{\vec{A}}}_{\mathrm{BG}})}{L(\hat{\sigma}_n^0,\hat{\vec{A}}_{\mathrm{BG}})}
\end{equation}
and the hats (\;$\hat{}$\;,\;$\hat{\hat{}}$\;) indicate that the parameters are those that maximise the extended likelihood $L$. Here $\vec{A}_{\mathrm{BG}}=\{A_{2 \nu \beta \beta}, A_{pp}, A_{\mathrm{Kr}}, A_{\mathrm{Rn}}, A_{\mathrm{Be}}, A_{\mathrm{mat}} \}$ are the amplitudes of the six background components discussed in section~\ref{sec:backgrounds}.

In our case the extended likelihood~\cite{Barlow:1990vc} (for a given value of the dark matter mass) is
\begin{equation}
\begin{split}
L(\sigma_n^0,\vec{A}_{\mathrm{BG}})&=\frac{\left(\muDM+\sum^6_{j=1} \muBGj\right)^N}{N!} \exp\left({-\muDM-\sum^6_{j=1} \muBGj} \right)\cdot \prod^6_{m=1} L_m(A_{\mathrm{BG}m})\\
&\cdot\prod^N_{i=1}\Biggl[\frac{\muDM}{\muDM+\sum^6_{k=1} \muBGk} \fDM(\Sone_i,\logSbS_i)\\
&\qquad+\sum^6_{j=1}\frac{\muBGj}{\muDM+\sum^6_{k=1} \muBGk } \fBGj (\Sone_i,\logSbS_i) \Biggr]\;,
\end{split}
\end{equation}
where $\muDM\propto \sigma_n^0$ and $\muBGj \propto A_{\mathrm{BG} j}$ are the mean number of events from dark matter and the background processes respectively, $\fDM$ and $\fBG$ are the unit normalised two-dimensional probability distribution functions for the signal and background processes in the $\Sone$ -- $\logSbS$ plane, $N$ is the total number of observed events and $\{\Sone_i,\logSbS_i\}$ are the values for a single event. Finally, $L_m(A_{\mathrm{BG}m})$ are the individual likelihood functions for the background normalisations, which we assume are Normal distributions with a standard deviation given by the respective error quoted in figure~\ref{fig:backrates}. As we are dealing with hypothetical experiments, we generate the unit normalised two-dimensional probability distribution functions $\fDM$ and $\fBGj$ from Monte Carlo by generating approximately two~million events for each process in the $\Sone$ -- $\logSbS$ plane. 

The results of Wilks~\cite{Wilks:1938dza} and Wald~\cite{Wald:1943} allow us to relate the significance with which we can reject the background-only hypothesis $(\sigma_n^0=0)$ to the test statistic in a simple way:
\begin{equation}
\label{eq:Z0}
Z_0=\sqrt{q_0}\;,
\end{equation}
where $Z_0$ is the number of standard deviations. In appendix~\ref{app:WilksWald}, we explicitly demonstrate that the approximation of Wald is good so that~eq.~\eqref{eq:Z0} is accurate. To obtain the discovery reach for each value of $\mDM$, we simulate a minimum of~2500 mock experiments and find the cross-section $\sigma_n^0$ for which 90\% of experiments have $Z_0 \geq3$. We will present a separate discovery reach for inelastic scattering off the~$\xea$ and~$\xeb$ isotopes. We are able to do this because, as figure~\ref{fig:StwoSone} shows, the two signal regions are well separated from each other and also from the nuclear recoil band, which contains events from the elastic scattering process.  The profile likelihood analysis takes into account the expected dark matter signal in the $\Sone$ -- $\logSbS$ plane so no cuts to identify a signal region are required. However in practice, to improve the run-time of our calculations, for each discovery reach calculation we restrict our analysis to the~$\Sone$ and~$\logSbS$ values around the dark matter signal region of interest. The restricted region is chosen to contain at least~$95\%$ of the events for each dark matter signal under consideration. By trying different regions, we found that our results are not sensitive to any reasonable choice. In appendix~\ref{app:cutcount}, we also provide a discovery reach calculation using a more conservative cut-and-count method. This serves as a useful cross-check against the profile likelihood analysis.

\section{Discovery reach for two-phase xenon detectors}
\label{sec:results}

\begin{figure}[t!]
\centering
\includegraphics[width=0.49\columnwidth]{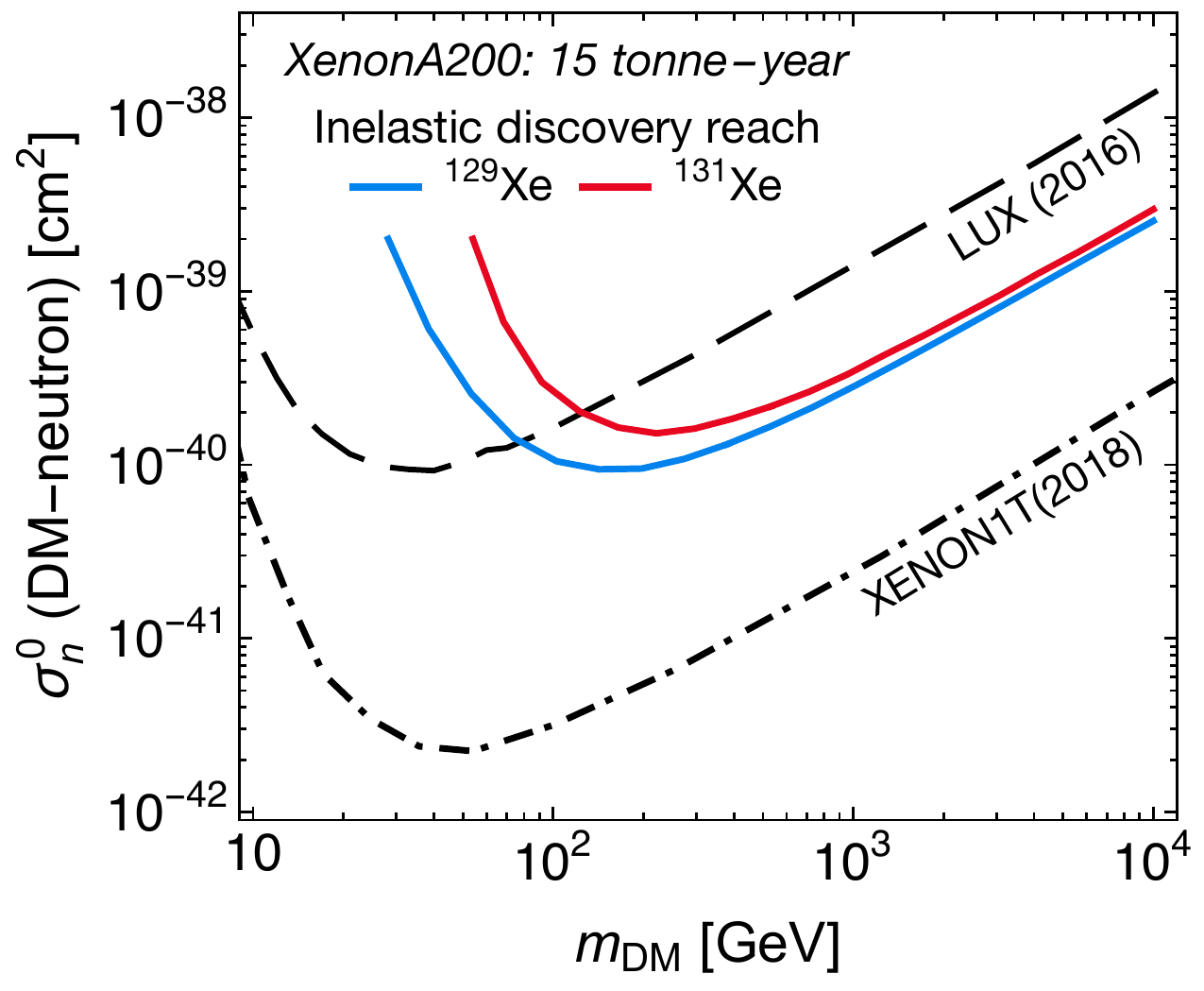} 
\includegraphics[width=0.49\columnwidth]{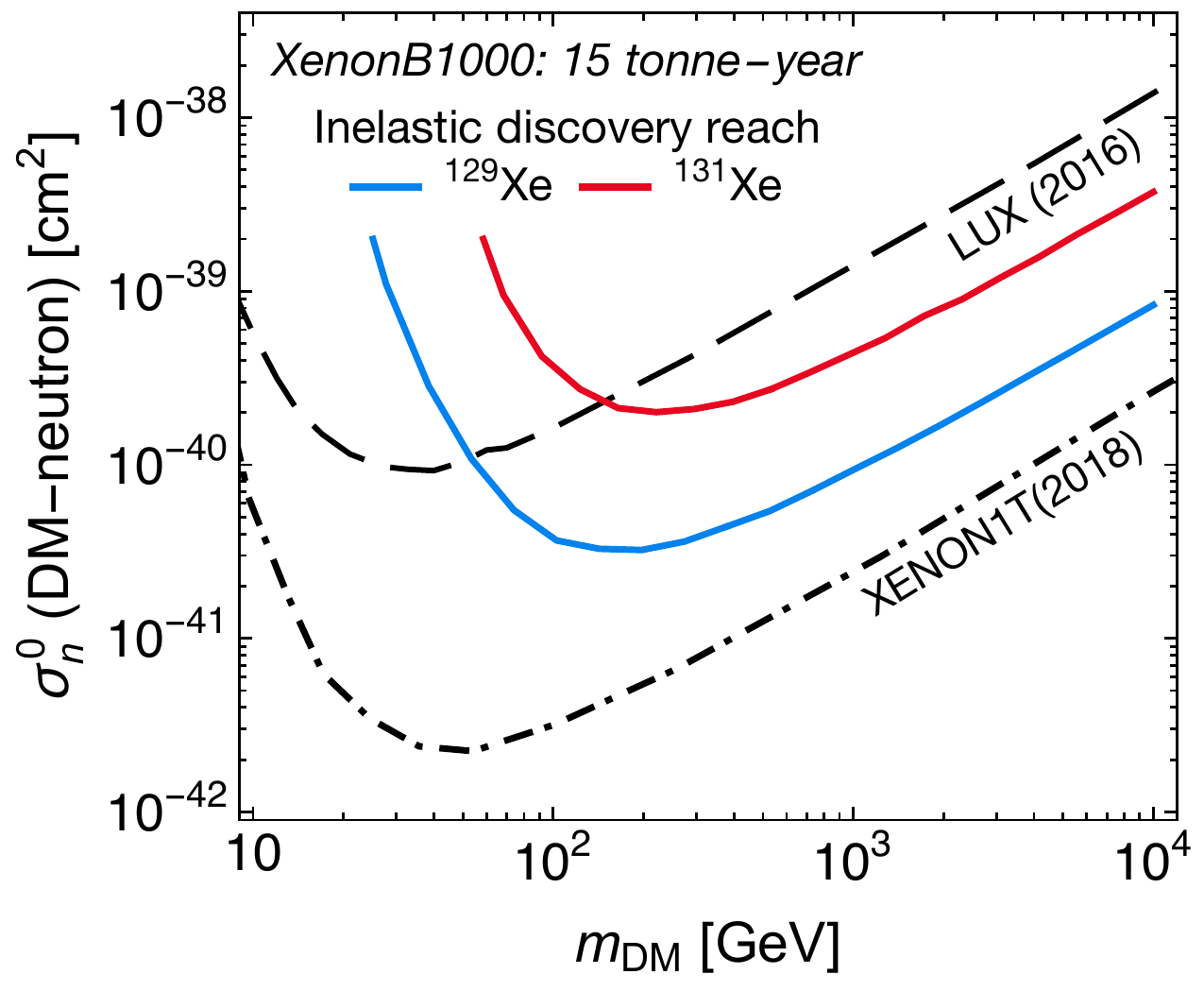}
 \caption{This figure shows the sensitivity of two-phase xenon experiments to the inelastic scattering process, which is our main result. The blue and red lines in both panels show the discovery reach for inelastically scattering off the~$\xea$ and~$\xeb$ isotopes, respectively. The left and right panels show the results for the {\it XenonA200} and {\it XenonB1000} benchmark scenarios assuming a 15~tonne-year exposure. In the parameter space above these lines, 90\% of experiments will make at least a~$3 \sigma$ detection of the inelastic signal. The discovery reach for inelastically scattering off the~$\xea$ isotope is better for the {\it XenonB1000} scenario, while for scattering off~$\xeb$, both benchmark scenarios have similar sensitivity.  Also shown is the LUX exclusion limit (black dashed line) from their search for elastically scattering dark matter and the projected exclusion limit from XENON1T assuming a two tonne-year exposure (black dot-dashed line).  \label{fig:limit}}
 \end{figure}

We present in figure~\ref{fig:limit} the main results of this paper. The red and blue lines show the discovery reach for detecting inelastic scattering off the~$\xea$ and~$\xeb$ isotopes for an exposure of 15~tonne-years. These lines show the smallest cross-section for which~90\% of experiments are able to make at least a~$3\sigma$ discovery of the inelastic signal. The left and right panels show the results for the {\it XenonA200} and {\it XenonB1000} benchmark scenarios (described in section~\ref{sec:detect}). The black dashed line shows the LUX 90\%~CL limit on the spin-dependent dark matter-neutron cross-section from their reanalysis of the 2013 search for dark matter that elastically scatters with the xenon isotopes~\cite{Akerib:2013tjd,Akerib:2015rjg,Akerib:2016lao}. The black dot-dashed line shows the projected limit from the XENON1T search for elastically scattering dark matter particles after an exposure of two tonne-years, which should be achieved by 2018. 

For both xenon isotopes and both benchmark scenarios, figure~\ref{fig:limit} shows that the discovery reach of the inelastic signal is below the current LUX exclusion limit for a dark matter mass greater than~$\sim100$~GeV. This means that for dark matter particles that are heavier than this, it is possible for the inelastic signal to be detected by a future two-phase xenon detector that collects an exposure of 15 tonne-year (such as LZ or XENONnT). The parameter space where the inelastic signal may be detected is populated by many dark matter models, including neutralino scenarios where the higgsino component is large (see e.g.~\cite{Cohen:2010gj,Chalons:2012xf,Bertone:2015tza}). XENON1T is expected to be significantly more sensitive than LUX and will probe all of the parameter space where the inelastic signal may be detected with a 15~tonne-year exposure. Therefore, if the inelastic signal is ever to be detected with this exposure, XENON1T should find evidence for the elastic scattering dark matter signal by~2018.

Comparing the left and right panels of figure~\ref{fig:limit}, we see that the discovery reach for inelastic scattering off the~$\xeb$ isotope (red line) is similar for both benchmark scenarios. This is because the ability to discriminate between signal and background processes is similar for both scenarios. In contrast, the discovery reach for inelastic scattering off the~$\xea$ isotope (blue line) is a factor~$\sim3.5$ lower for the {\it XenonB1000} benchmark scenario. This is because, as figure~\ref{fig:StwoSone} shows, more of the signal region lies below the electronic recoil band for the {\it XenonB1000} scenario so the discrimination power is better (cf.~the discussion at the end of section~\ref{subsec:comp} where the discrimination power was quantified).

 \begin{figure}[t!]
\centering
\includegraphics[width=0.49\columnwidth]{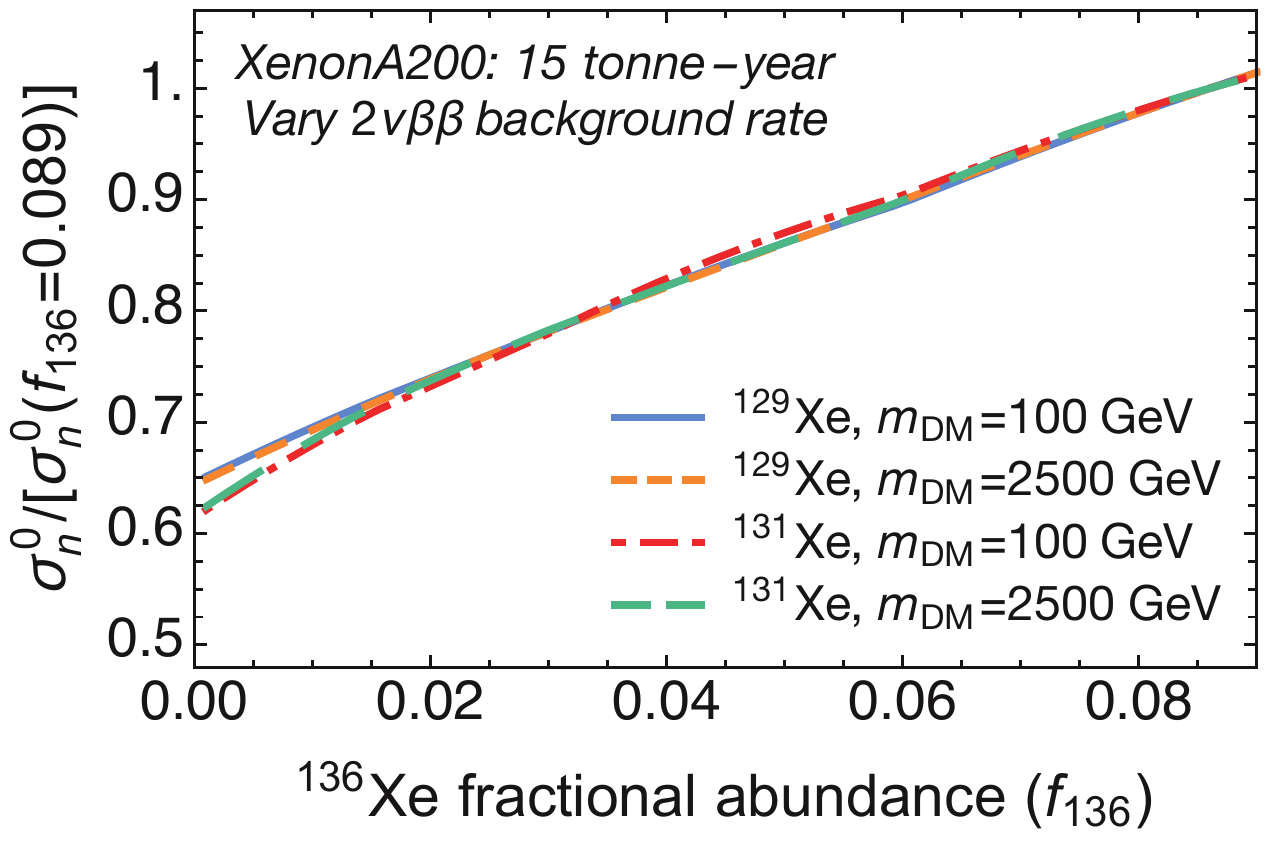} 
\includegraphics[width=0.49\columnwidth]{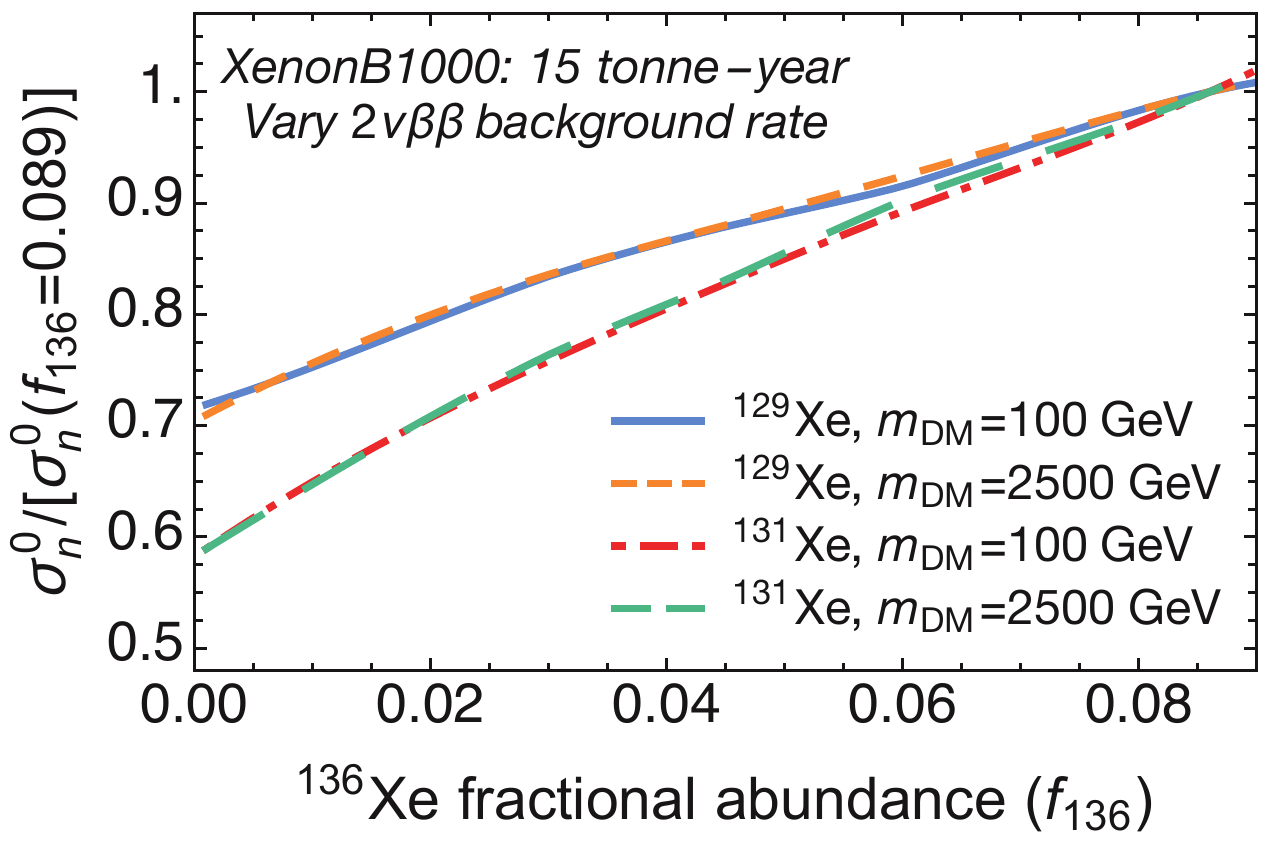} \\ \vspace{3mm}
\includegraphics[width=0.49\columnwidth]{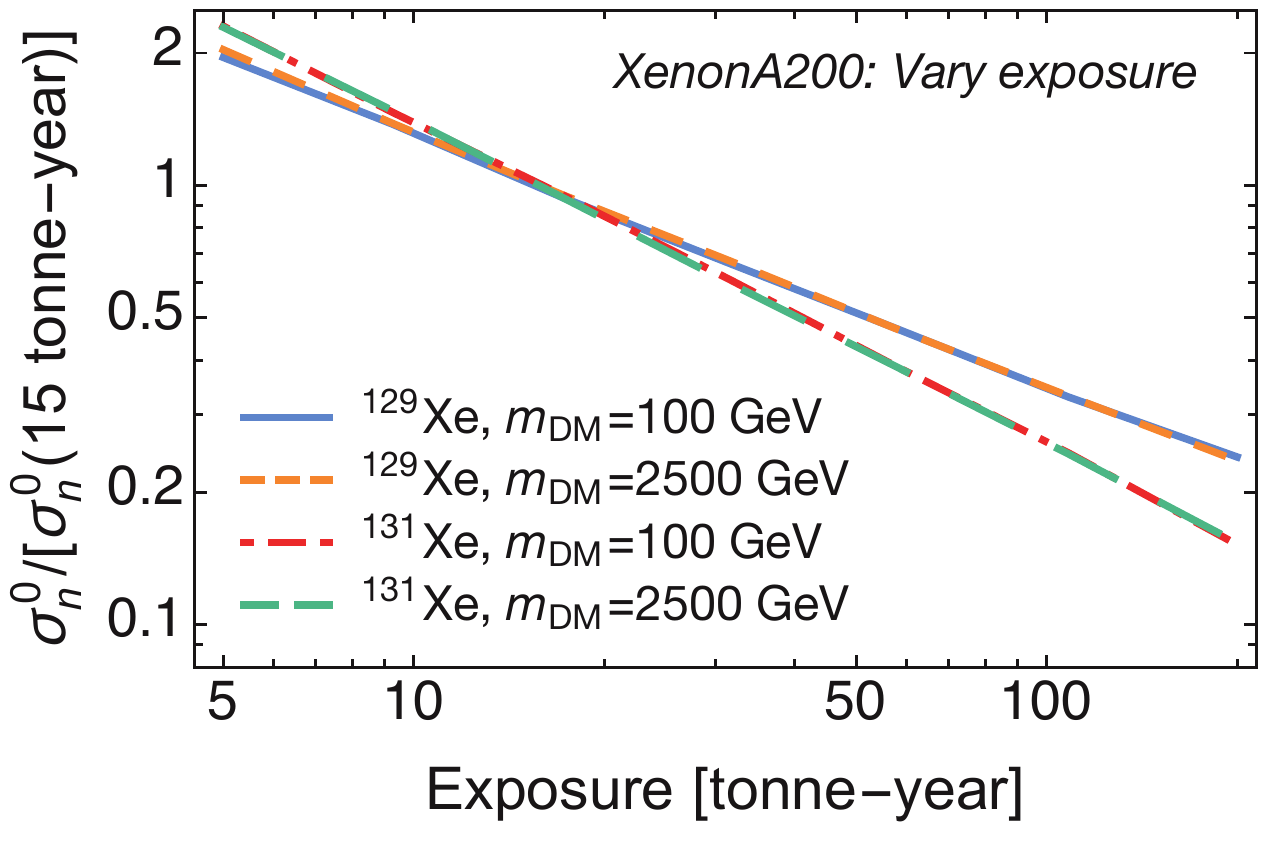} 
\includegraphics[width=0.49\columnwidth]{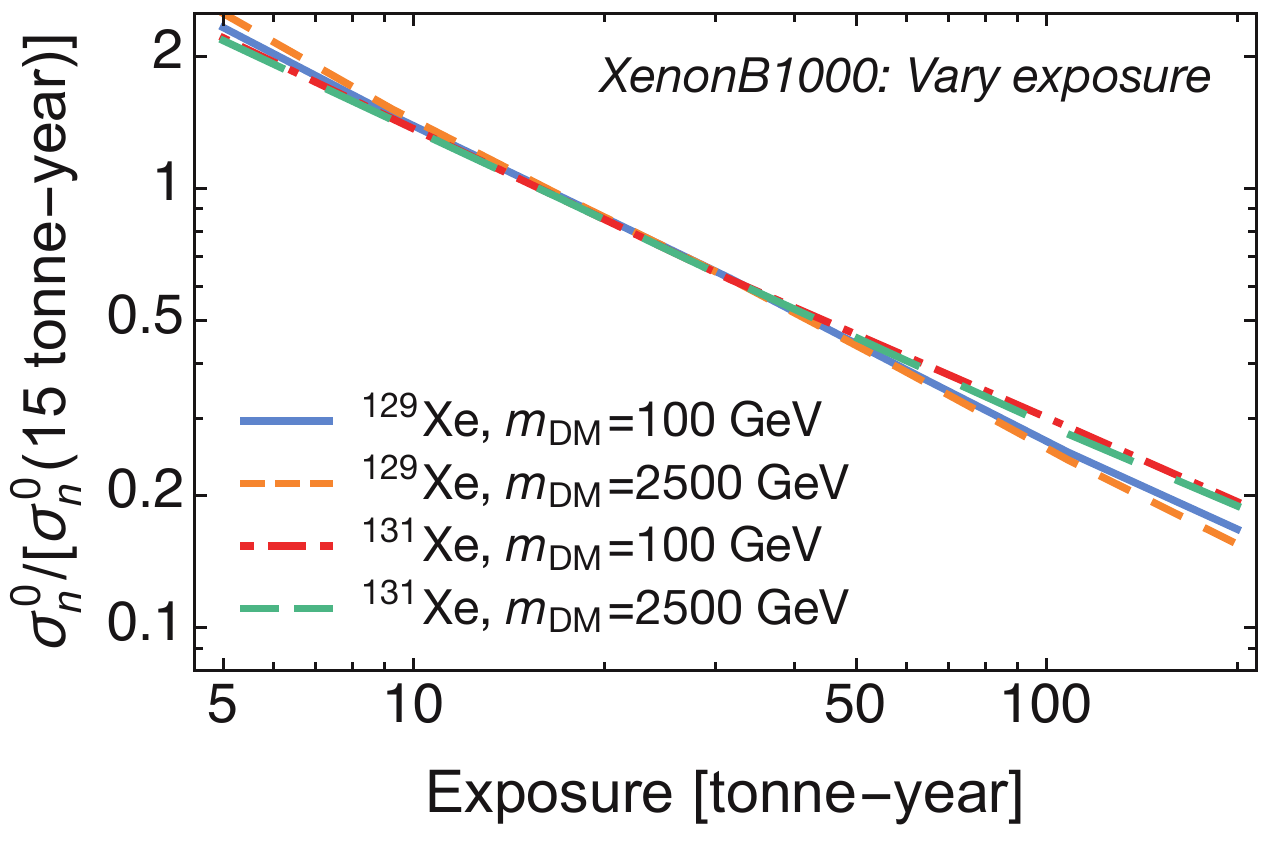} 
 \caption{The upper panels show how the discovery reach changes as the rate of the main background, the $2\nu\beta\beta$-decay of the $\xec$ isotope, is varied. The cross-section is normalised to the discovery reach for an abundance of~$8.86\%$, the~$\xec$ abundance of natural xenon. The lower panels show the variation in the discovery reach for different exposures, normalised to the cross-section for a 15~tonne-year exposure.  In each panel, we show the discovery reach for two values of the dark matter mass and find that the variation does is independent of the mass. The left and right panels show the result for our two benchmark scenarios. They show that the variation is only weakly dependent on the detector parameters. By reducing the~$\xec$ abundance to 1\%, the same sensitivity can be achieved with an exposure that is $\sim35\%$ smaller. The inelastic signal search regions are not background free so the discovery reach scales only as $\sim(\mathrm{exposure})^{-0.7}$. \label{fig:variations}}
 \end{figure}

The discovery reach shown in figure~\ref{fig:limit} assumes an exposure of 15 tonne-years and the backgrounds rates discussed in section~\ref{sec:backgrounds}. We now explore how the discovery reach changes as we vary these assumptions. Firstly, we examine variations in the background rate. As we showed in section~\ref{sec:backgrounds}, the dominant background is the from the $2\nu\beta\beta$-decay of the $\xec$ isotope so it is possible to reduce this background rate by reducing the abundance of the~$\xec$ isotope. Depleting (or enriching) the~$\xec$ isotope from xenon is relatively straightforward as demonstrated by experiments that use xenon enriched in~$\xec$ to search for neutrinoless double beta decay. The upper two panels of figure~\ref{fig:variations} show how the discovery reach changes as we vary the~$\xec$ abundance. The left and right panels show the results for the {\it XenonA200} and {\it XenonB1000} benchmark scenarios, respectively. The results are shown for two dark matter mass values and we plot the discovery reach cross-section normalised to the discovery reach assuming that the fractional abundance of~$\xec$ is~$8.86\%$, which is the abundance in natural xenon and is the value that we have assumed throughout the paper. As expected, the discovery reach extends to smaller values of the cross-section as the fractional abundance is reduced.  Figure~\ref{fig:variations} shows that the variation does not depend on the dark matter mass and is only weakly dependent on the benchmark scenario. For both scenarios, we see that lowering the~$\xec$ abundance to~$1\%$ means that the smallest cross-section for which the inelastic signal may be discovered is reduced by $\sim35\%$ . 

Secondly, we examine variations in the exposure. These results are shown in the lower two panels of figure~\ref{fig:variations}, where the discovery reach cross-section has been normalised to the discovery reach assuming a 15~tonne-year exposure.  Figure~\ref{fig:variations} again demonstrates that the variation does not depend on the dark matter mass and is only weakly dependent on the benchmark scenario. The discovery reach for inelastically scattering off~$\xea$ for the {\it XenonA200} scenario decreases more slowly as the exposure increases compared to the other scenarios because this signal region is most dominated by background processes (cf.~the discussion at the end of section~\ref{subsec:comp}). For a background free signal region, the improvement in the discovery reach is expected to scale as $(\mathrm{exposure})^{-1}$. As there is always some background contamination for the inelastic signals and the benchmark scenarios that we consider, we instead find that the discovery reach scales as $\sim(\mathrm{exposure})^{-0.7}$. In practice, this means that for a 200 tonne-year exposure, a benchmark exposure used in sensitivity studies for DARWIN, the various discovery reach lines presented in figure~\ref{fig:limit} should be lowered by a factor~$\sim 5$. 

We end this section by returning to the question of whether a two tonne-year exposure of XENON1T can probe all of the parameter space where the inelastic signal may be discovered. With the scaling of the exposure determined in figure~\ref{fig:variations}, we find that the exposure required to reach the XENON1T exclusion limit for all scenarios except the~$\xea$ signal in the {\it XenonB1000} benchmark scenario is~$\sim 500$~tonne-year. Such a large exposure is unlikely to be achieved in the foreseeable future. For the~$\xea$ signal in the {\it XenonB1000} scenario, an exposure of approximately $\{225, 90, 70 \}$~tonne-year for $\mDM=\{150, 1000, 10000 \}~\mathrm{GeV}$ is required for the discovery cross-section to reach the XENON1T limit shown in figure~\ref{fig:limit}. The exposure can be reduced to approximately $\{165, 60, 45 \}$~tonne-year if the~$\xec$ abundance is reduced to~$1\%$. This demonstrates that it may be possible to discover inelastic scattering off the~$\xea$ isotope even for cross-sections below the XENON1T limit, but only for optimal detector parameters (as in the {\it XenonB1000} scenario) and with large exposures that will only be achieved with detectors such as DARWIN.

\section{Conclusions and outlook}
\label{sec:con}

The canonical search for dark matter with direct detection experiments is for an elastic scattering process where the dark matter simply causes the nucleus to recoil. It was long ago realised that low-lying inelastic transitions of the nucleus may also play an important role since the dark matter's kinetic energy is sufficient to excite the target nucleus. In this instance, rather than just measuring the recoil of the nucleus, direct detection experiments measure the nuclear recoil energy together with the photon-energy released when the nucleus transitions back to the ground state (see figure~\ref{fig:scattering}). 

The inelastic scattering rate does not have the nucleon-number--squared enhancement~$(\sim10^4)$ found with elastic {\it spin-independent} interactions so the inelastic signal will only be measurable for {\it spin-dependent} interactions, whose elastic scattering rate also does not have the nucleon-number--squared enhancement. Two-phase xenon detectors are an excellent probe of the elastic and inelastic spin-dependent interaction having two isotopes sensitive to these processes,~$\xea$ and~$\xeb$, that each comprise approximately~$25\%$ of natural xenon and have low-lying excitations at $39.6~\mathrm{keV}$ and $80.2~\mathrm{keV}$, respectively. The purpose of this paper was to quantify the sensitivity of future tonne-scale two-phase xenon experiments, such as LZ, XENONnT and DARWIN, to the inelastic signal. We do this for the axial-vector interaction (eq.~\ref{eq:A-V}), for which accurate calculations of the nuclear structure functions are available.

We considered two benchmark scenarios, {\it XenonA200} and {\it XenonB1000} (described in section~\ref{sec:detect}), whose most important difference is the applied drift field of $200~\mathrm{V/cm}$ and $1000~\mathrm{V/cm}$, respectively. The parameters in these scenarios were chosen because they should bracket the performance of future experiments. We implemented a realistic Monte Carlo simulation of a two-phase xenon detector to model these scenarios, relying on the NEST phenomenological model to describe the interactions of the nucleus and photon in liquid xenon. This was vital so that we could translate energies into the measurable quantities, the primary (S1) and secondary (S2) scintillation signals (see figure~\ref{fig:S1S2plot}). We also had to quantify the background rates, finding that the $2\nu\beta\beta$-decay of~$\xec$ dominates (see figure~\ref{fig:backrates}).

We demonstrated that two-phase xenon detectors allow for some discrimination between the inelastic signal and the background events because the signal region has a smaller $\log_{10}\left(\Stwob/\Sone \right)$ value compared to the main backgrounds (see figure~\ref{fig:StwoSone}). Our main results were shown in figures~\ref{fig:limit} and~\ref{fig:variations}, where we quantified the sensitivity of our benchmark scenarios to the inelastic signal in terms of the discovery reach, which is the smallest cross-section for which~90\% of experiments detect the signal with at least a~$3\sigma$ significance.  This cross-section is below the current LUX exclusion limit for a dark matter mass greater than~$\sim100$~GeV, implying that for dark matter particles that are heavier than this, it is possible for the inelastic signal to be detected with a future two-phase xenon detector. Except in the case of optimal detector parameters (as in the {\it XenonB1000} scenario) and large exposures (more than 50~tonne-years), XENON1T, with a two tonne-year exposure, will probe all of the parameter space where the inelastic signal may be detected with their search for elastically scattering dark matter. 

We end by discussing some of the possible extensions of this work. Firstly, we were only able to consider the inelastic signal from the axial-vector interaction since this is the only spin-dependent interaction for which the inelastic structure functions have been calculated. It would be desirable to calculate the discovery reach of other spin-dependent operators such as VA $(\bar{\chi}\gamma^{\mu}\chi \bar{\psi}_q\gamma_{\mu} \gamma^5\psi_q)$ or SP $(\bar{\chi}\chi  \bar{\psi}_q \gamma^5\psi_q)$. Secondly, in this work we assumed that nuclear recoil and photon scintillation signals could not be distinguished. It may be possible that for some fraction of the events, the photon travels sufficiently far from the initial interaction to give a distinctive pulse shape different from background events. This would further improve the sensitivity to inelastic scattering signals. Thirdly, we focussed solely on two-phase xenon experiments as these are better able to distinguish between signal and background signals compared to single-phase xenon detectors. However it may be possible that tonne-scale single-phase detectors can improve their sensitivity to the inelastic signal by performing an annual modulation search or by modelling the~$\Sone$ pulse shape. Finally, we have stated that a detection of the inelastic signal together with the elastic signal would point strongly to a spin-dependent interaction over a spin-independent interaction from a single xenon experiment. It would be desirable to have a dedicated study to quantify this statement and to concretely demonstrate how a measurement of both signals would help pin down the nature of the interaction between dark matter and the Standard Model particles.

\acknowledgments{
I am particularly grateful to Nassim Bozorgnia for discussions that reignited my interest in this topic and for providing data from the EAGLE simulation, and to Andrew Brown for answering {\it many} of my naive questions. I'm also grateful for discussions with Rafael Lang and Simon Fiorucci at the Rencontres de Blois conference, with Martin Hoferichter, Philipp Klos and Achim Schwenk at the Mainz Institute for Theoretical Physics (MITP), with Alastair Currie and Marc Schumann at the TAUP conference, to Felix Kahlhoefer for comments on an early draft of this manuscript, and to Sebastian Liem for discussions on statistics. This work is part of the research programme of the Foundation for Fundamental Research on Matter~(FOM), which is part of the Netherlands Organisation for Scientific Research~(NWO).
}

\appendix

\section{Mean photon and electron yields}
\label{app:meanyields}

We use NEST's semi-empirical model to calculate the mean number of photons and electrons from nuclear and electromagnetic interactions. In order that our results can be easily reproduced, in this appendix we collect the parameters and formulae that we use. The energy is denoted $E$ and has units keV while the applied electric field is denoted by $F$ and has units $\mathrm{V}/\mathrm{cm}$.

\subsection{Nuclear recoils}

Our treatment of nuclear recoils follows~\cite{Lenardo:2014cva}. For an energy deposit $E$, the number of quanta $\nquanta$ is
\begin{equation}
\nquanta=\frac{L E}{W}\;,
\label{app:nq}
\end{equation}
where $\nquanta=\nion+\nex$ is the total number of ions $\nion$ and excitons $\nex$ respectively and $W=13.7~\mathrm{eV}$ is the average energy to produce a quanta. The quenching factor (to account for energy lost to atomic motion) is 
\begin{equation}
L=\frac{k g(\epsilon)}{1+k g(\epsilon)}\;,
\end{equation}
where $k=0.1394$, $\epsilon=11.5 E Z^{-7/3}$ ($Z=54$ for xenon) and $g(\epsilon)=3 \epsilon^{0.15}+0.7 \epsilon^{0.6}+\epsilon $.

The number of ions and excitons follows the ratio
\begin{equation}
\nex/\nion=\alpha F^{-\zeta}(1-e^{\beta \epsilon})\;,
\end{equation}
where $\zeta=0.0472$, $\alpha=1.240$ and $\beta=239$. The probability that an ion recombines is
\begin{equation}
r=1-\frac{\ln(1+\nion \varrho)}{\nion \varrho}\;,
\end{equation}
where $\varrho=\gamma F^{-\delta}$, $\gamma=0.01385$ and $\delta=0.0620$.

The number of electrons $\nele$ and photons $\ngam$ is given by
\begin{align}
\nele&=\nion-r\nion\\
\ngam&=f_{l}(\nex+r\nion)\;,
\end{align}
where
\begin{equation}
f_l=\frac{1}{1+\eta e^{\lambda}}\;,
\end{equation}
is another quenching factor (accounting for the Penning effects, when two excitons interact to produce one exciton and one photon), $\eta=3.3$ and $\lambda=1.14$.

\subsection{Electronic recoils from incident beta particles}

The equivalent formulae for electronic recoils are generally simpler since there are no quenching factors. Our treatment follows~\cite{NESTbench}. The number of quanta is
\begin{equation}
\nquanta=\frac{E}{W}\;,
\end{equation}
where $W$ is the same as in eq.~\eqref{app:nq}, the ratio of excitons to ions is
\begin{equation}
\nex/\nion=0.15\;,
\end{equation}
the probability that an ion recombines is
\begin{equation}
r=1-\frac{\ln(1+\nion \tilde{\varrho})}{\nion \tilde{\varrho}}
\end{equation}
and the number of electrons $\nele$ and photons $\ngam$ is
\begin{align}
\label{eq:Ane1}
\nele&=\nion-r\nion\\
\label{eq:Ane2}
\ngam&=\nex+r\nion\;.
\end{align}
The expression for $\tilde{\varrho}$ is more complicated in this case, depending on both~$E$ and~$F$. We have that
\begin{equation}
\tilde{\varrho}=\frac{\mathfrak{F}}{4 E^{-\mathfrak{E}}}\left\{1-\exp{\left[-\left(\frac{E-\mathfrak{Z}}{\mathfrak{A}} \right)^{0.188 F^{1/3}}\right]} \right\}\;,
\end{equation}
where
\begin{align}
\mathfrak{F}&=0.6347 \exp{\left(-1.4\times10^{-4} F\right)}\\
\mathfrak{E}&=1.5-0.373 \exp{\left(- 10^{-3} F/ \mathfrak{F}\right)}\\
\mathfrak{A}&=10 F^{-0.04} \exp{\left(18/F \right)}\\
\mathfrak{Z}&=4-F^{0.2147}\;.
\end{align}

\subsection{Electronic recoils from incident gamma particles}

\begin{table}[t]
\setlength{\tabcolsep}{5pt}
\renewcommand{\arraystretch}{1.3}
\center
\begin{tabular}{c|cc} 
$F~[\mathrm{V/cm}]$ & $\alpha_{\gamma}(39.6~\mathrm{keV}, F)$  & $\alpha_{\gamma}(80.2~\mathrm{keV}, F)$\\
\hline
200 &0.11  &0.18 \\
1000 & 0.21 &0.24
\end{tabular}
\caption{The rescaling $\alpha_{\gamma}$ factor to convert between the yields for incident beta particles and incident gamma particles. We give the values for the two de-excitation energies and for the values of the drift fields that we consider in this study.}
\label{tab:beta2gammafactors}
\end{table}

There is a small difference between the electron and photon yields from incident gamma and beta particles~\cite{Szydagis:2011tk}. The difference arises because an incident gamma produces multiple lower energy electron recoils. For instance, a 39.6 keV gamma typically results in Auger and photo-electrons with energies 4.5~keV, 4.8~keV, 5.3~keV and 25~keV respectively~\cite{Dahl:2009nta}. A complete description of these events requires a full simulation with NEST, which tracks the energy deposition of the incident gamma. Such a simulation is beyond the scope of this work. Instead we use the mean yields from a NEST simulation presented in fig.~1 of~\cite{Szydagis:2013sih} to rescale the yields from an incident beta particle:
\begin{align}
n_{\gamma}\left[\gamma(E,F) \right]&=\left(1+\alpha_{\gamma}(E,F) \right) n_{\gamma}\left[\beta(E,F) \right]\;.
\end{align}
Here $n_{\gamma}\left[\beta(E,F) \right]$ is the yield from a beta-particle, given by eq.~\eqref{eq:Ane2}. We have made it explicit that the yields depend on both the incident energy~$E$ and drift field~$F$. This rescaling is sufficient to also calculate the change in the number of electrons. Quanta conservation tells us that $n_e=n_{\rm{quanta}}-n_{\gamma}$ (this follows from the usual assumption that energy lost to heat can be ignored for electromagnetic interactions), so increasing $n_{\gamma}$ for incident gamma particles automatically decreases $n_e$, as shown in~\cite{Szydagis:2013sih}.  We only require the rescaling $\alpha$ factors at the de-excitation energies: $39.6~\mathrm{keV}$ and $80.2~\mathrm{keV}$. In table~\ref{tab:beta2gammafactors} we give the factors at these energies and at the two values of the drift field that we consider in this study. 

\section{Testing the approximations of Wilks and Wald}
\label{app:WilksWald}

\begin{figure}[t!]
\centering
\includegraphics[width=0.48\columnwidth]{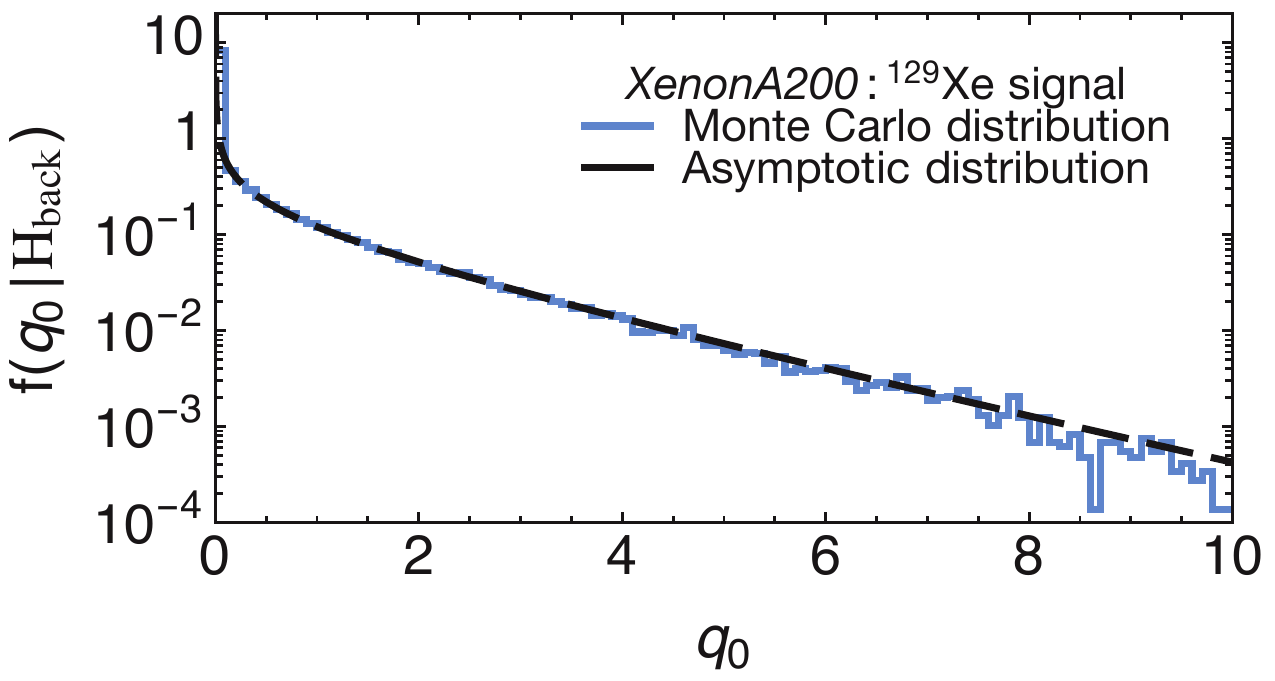} 
\includegraphics[width=0.48\columnwidth]{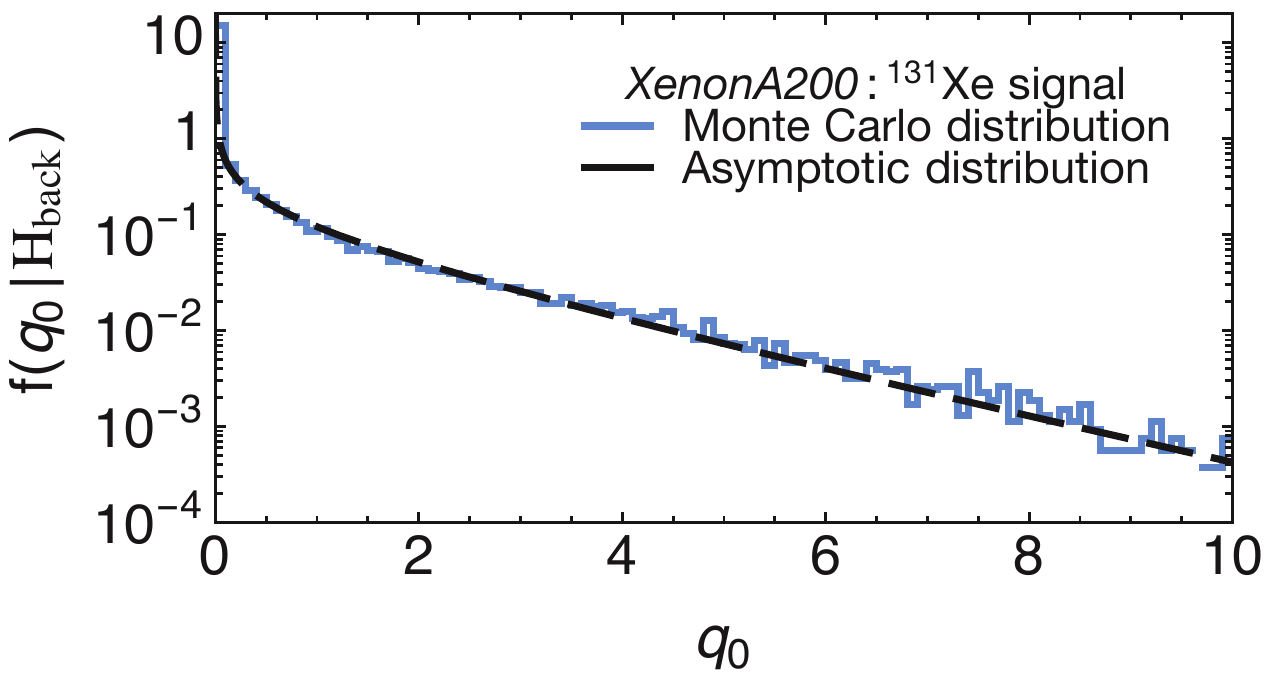}\\
\includegraphics[width=0.48\columnwidth]{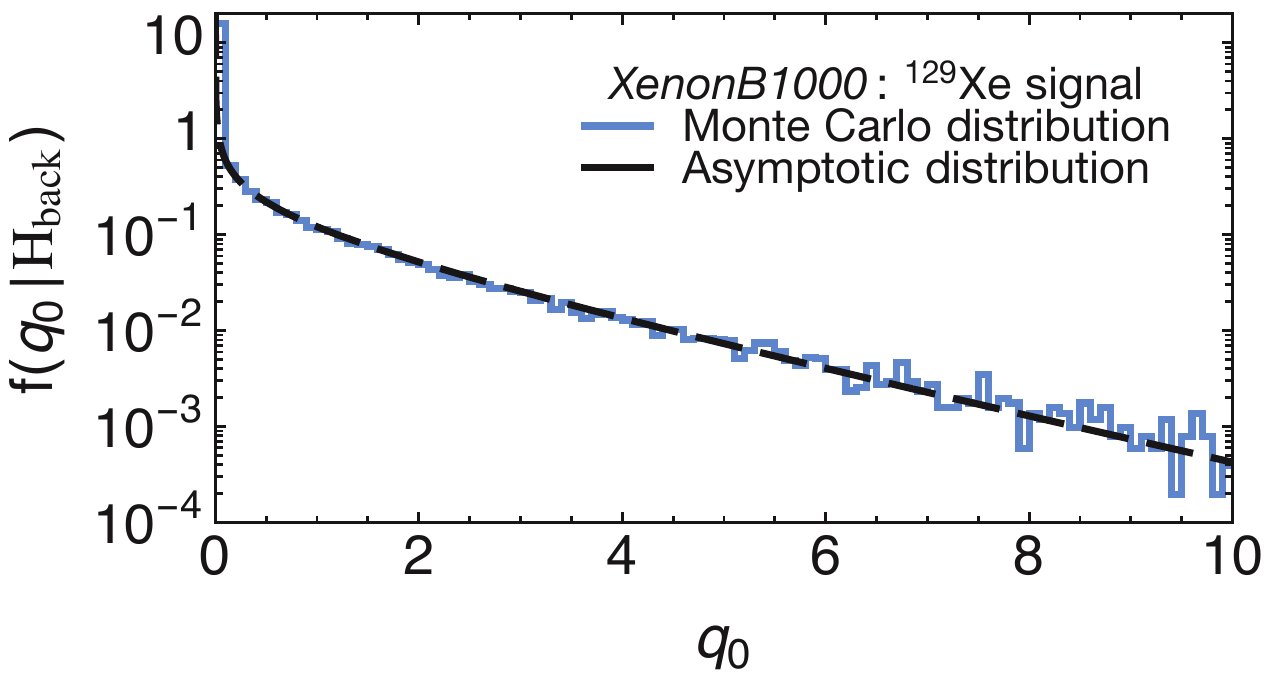}
\includegraphics[width=0.48\columnwidth]{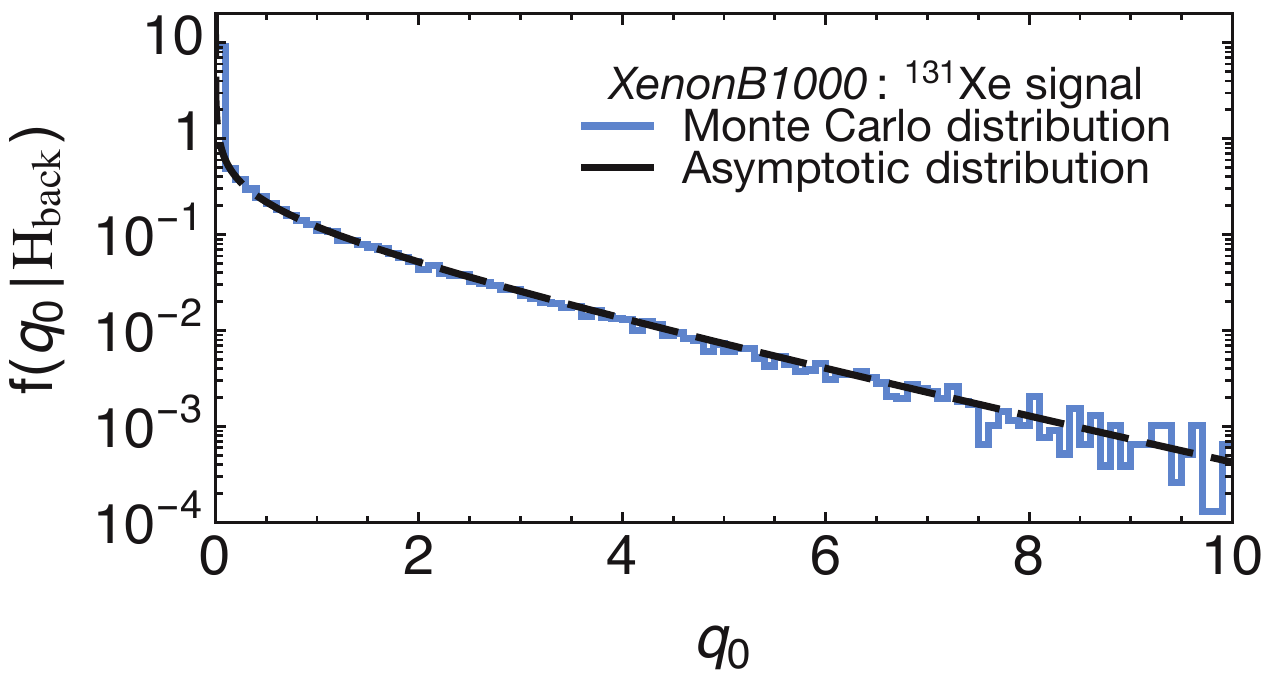}
 \caption{A comparison of the distribution of the test statistic $q_0$ under the background-only hypothesis from a Monte Carlo simulation (blue histogram) and from the asymptotic formula that follows from Wald and Wilks (black dashed line). The four panels show the result for the four signal regions that we consider: the signal regions from scattering off the~$\xea$ and~$\xeb$ isotopes for the {\it XenonA200} and {\it XenonB1000} benchmark scenarios. The Monte Carlo and asymptotic formula agree well, with the implication that our use of eq.~\eqref{eq:Z0} is robust. \label{fig:wald}}
 \end{figure}

As discussed in~\cite{Cowan:2010js}, if Wald's approximation is good, then following the result of Wilks, the distribution of the test statistic $q_0$ under the background-only hypothesis $\mathrm{H}_{\rm{back}}$ should asymptotically follow
\begin{equation}
\label{eq:wald}
f(q_0|\mathrm{H}_{\mathrm{back}})=\frac{1}{2} \delta(q_0)+\frac{1}{2} \frac{1}{\sqrt{2 \pi q_0}}e^{-q_0/2}\;.
\end{equation}
In figure~\ref{fig:wald} we demonstrate that Wald's approximation is good by comparing $f(q_0|\mathrm{H}_{\mathrm{back}})$ from~100000 Monte Carlo simulations (blue histogram) with~eq.~\eqref{eq:wald} (black dashed line). We find good agreement between the asymptotic distribution and our Monte Carlo distribution for the four signal regions of interest. This implies that our use of eq.~\eqref{eq:Z0} is justified.

\section{A cut-and-count analysis}
\label{app:cutcount}

\begin{table}[t]
\centering
\begin{tabular}{@{}l c c c c  @{}}\toprule 
& \multicolumn{2}{c}{{\it XenonA200}\,: $^{129}\mathrm{Xe}$} & \multicolumn{2}{c}{{\it XenonA200}\,: $^{131}\mathrm{Xe}$} \\
 & $\Sone$ only &  $\Sone$ \& $\logSbS$ & $\Sone$ only &  $\Sone$ \& $\logSbS$ \\ \midrule
$2\nu \beta\beta$ & 7483 & $15.5$ & $22767$ & 27.2 \\
$pp$& 1673& $6.0$& $1687$ & 4.1 \\ 
$^7\mathrm{Be}$ & 170& $0.6$& $250$ & 0.4  \\
$^{85}\mathrm{Kr}$& 1066& $3.4$ & $1609$ & 3.0  \\
$^{222}\mathrm{Rn}$& 235& $0.7$& $360$ & 0.7  \\
Materials& 38& $0.1$& $56$ & 0.1  \\ \midrule
BG total  & 10665 & 26.3& 26729 & 35.5  \\ \midrule
DM total & 113 & $25.6$ & $38$ & 18.7  \\  \bottomrule
\toprule 
& \multicolumn{2}{c}{{\it XenonB1000}\,: $^{129}\mathrm{Xe}$} & \multicolumn{2}{c}{{\it XenonB1000}\,: $^{131}\mathrm{Xe}$} \\
 & $\Sone$ only &  $\Sone$ \& $\logSbS$ & $\Sone$ only &  $\Sone$ \& $\logSbS$ \\ \midrule
$2\nu \beta\beta$ & 16659 & $8.5$ & $47901$ & 31.8 \\
$pp$& 2316& $3.4$& $1743$ & 4.7 \\ 
$^7\mathrm{Be}$ & 268& $0.3$& $368$ & 0.5  \\
$^{85}\mathrm{Kr}$& 1699& $1.9$ & $2359$ & 3.4  \\
$^{222}\mathrm{Rn}$& 383& $0.4$& $1004$ & 0.8  \\
Materials& 69& $0.1$& $779$ & 0.1  \\ \midrule
BG total  & 21394 & 14.6& 54154 & 41.3  \\ \midrule
DM total & 112 & $52.7$ & $38$ & 17.2  \\  \bottomrule
\end{tabular} 
\caption{The upper and lower tables give the number of background and signal events for a 15~tonne-year exposure for the {\it XenonA200} and {\it XenonB1000} benchmark scenarios respectively. The different columns show the number of events in the different signal regions employed. The details of each signal region is given in the text of appendix~\ref{app:cutcount}. The number of dark matter events assume $\mDM=1000~\mathrm{GeV}$ and $\sigma_n^0=5\times 10^{-40}~\mathrm{cm}^2$.
\label{tab:cut}}
\end{table}

In the calculation of the discovery limits in the main body of the paper, we employed a profile likelihood analysis that takes into account the position of each event in the $\Sone$ -- $\logSbS$ plane. In this appendix, we present an alternative calculation of the discovery limit using a more conservative cut-and-count analysis. In this case, we define a signal box in the $\Sone$ -- $\logSbS$ plane and simply count the number of events that fall in this signal region.

The number of background and signal events for a 15~tonne-year exposure for each of the benchmark scenarios and for the two xenon isotopes are shown in table~\ref{tab:cut}. The upper and lower tables are for the {\it XenonA200} and {\it XenonB1000} benchmarks scenarios respectively. As explained in section~\ref{sec:backgrounds}, the dominant background is always from the $2\nu\beta\beta$-decay of $\xec$. The number of dark matter events quoted are for a dark matter particle with mass $\mDM=1000~\mathrm{GeV}$ and a cross-section $\sigma_n^0=5\times 10^{-40}~\mathrm{cm}^2$.

The various signal regions in the $\Sone$ -- $\logSbS$ plane were taken as follows:

\begin{itemize}
\item {\it XenonA200}, $\xea$: `$\Sone$ only' refers to the range $125~\mathrm{PE}\leq\Sone\leq275~\mathrm{PE}$. `$\Sone$~\& $\logSbS$' has the additional constraint $1.15\leq\logSbS\leq1.36$.

\item {\it XenonA200}, $\xeb$: `$\Sone$ only' refers to the range $260~\mathrm{PE}\leq\Sone\leq475~\mathrm{PE}$. `$\Sone$~\& $\logSbS$' has the additional constraint $1.1\leq\logSbS\leq1.4$.

\item {\it XenonB1000}, $\xea$: `$\Sone$ only' refers to the range $180~\mathrm{PE}\leq\Sone\leq420~\mathrm{PE}$. `$\Sone$~\& $\logSbS$' has the additional constraint $1.7\leq\logSbS\leq2.0$.

\item {\it XenonB1000}, $\xeb$: `$\Sone$ only' refers to the range $350~\mathrm{PE}\leq\Sone\leq650~\mathrm{PE}$. `$\Sone$~\& $\logSbS$' has the additional constraint $1.8\leq\logSbS\leq2.1$.

\end{itemize}

These signal regions were chosen `by eye' with reference to figure~\ref{fig:StwoSone} so it may be possible that this calculation could be improved by choosing more optimal signal regions. As this calculation mainly serves as a cross-check of the calculation in the main body of the paper, we have not investigated this further.

\begin{figure}[t!]
\centering
\includegraphics[width=0.49\columnwidth]{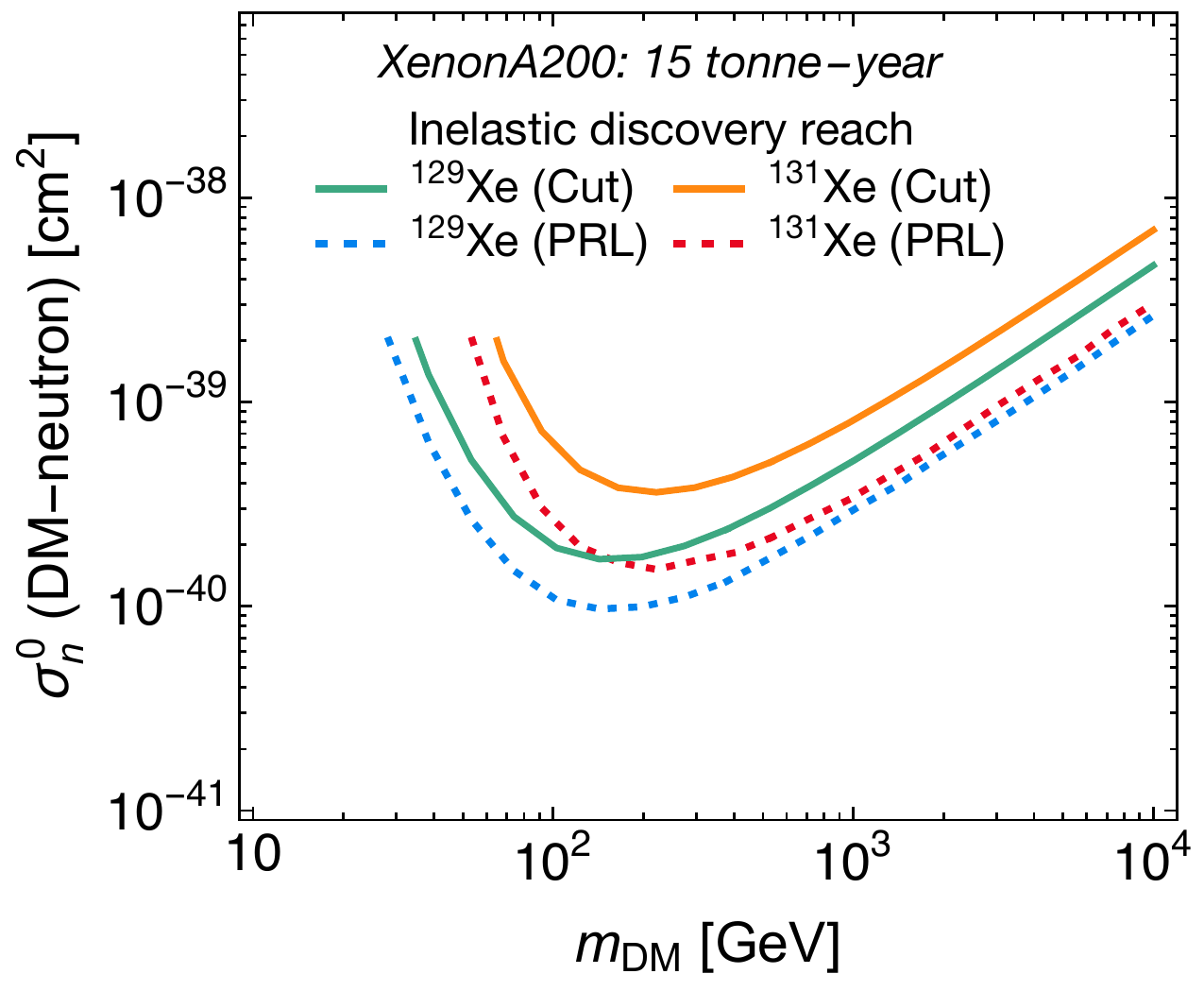} 
\includegraphics[width=0.49\columnwidth]{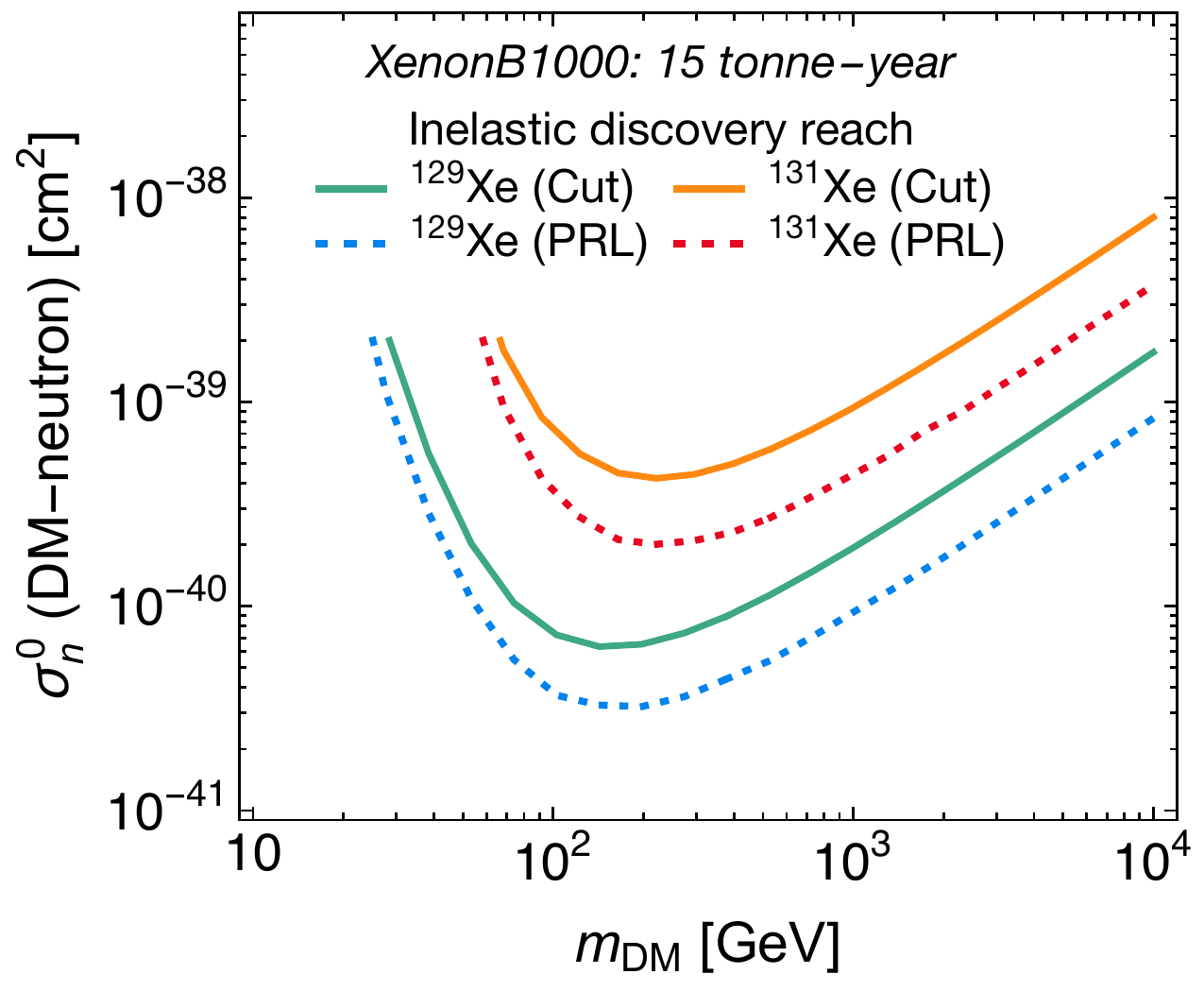}
 \caption{ A comparison of the discovery reach calculations using the cut-and-count approach discussed in this appendix (solid green and orange line) and the approach taken in the main body of the paper that takes into account the position of each event in the $\Sone$ -- $\logSbS$ plane. The two calculations give comparable results, with the cut-and-count approach giving a limit on $\sigma_n^0$ that is higher since it uses less of the information measured by our mock experiments.
   \label{fig:cutlimit}}
 \end{figure}

In this calculation, we again employ the profile likelihood ratio described in section~\ref{sec:discovery}. However in this instance, the extended likelihood is
\begin{equation}
L(\sigma_n^0,A_{\mathrm{BG}})= \frac{\left(\muDM+\muBG\right)^N}{N!} \exp \left(-\muDM- \muBG \right)  L(A_{\mathrm{BG}})\;.
\label{eq:Lcut}
\end{equation}
We make a further simplifying assumption and consider only one normalisation of the background, rather than having an individual normalisation for each background component. In eq.~\eqref{eq:Lcut},  $L(A_{\mathrm{BG}})$ is a Normal distribution that accounts for the uncertainty of the background, which we assume has an error of $4\%$ (obtained by combining the uncertainties from the individual components).

The discovery limits that follows from this calculation are shown by the solid green and orange lines in figure~\ref{fig:cutlimit}. Also show by the red and blue dashed lines are the discovery limits from the main body of the paper. As we would naively expect, the cut-and-count discovery limit on $\sigma_n^0$ is higher than the result in the main body of the paper. This is because the cut-and-count calculation uses less of the information measured by our mock experiments. However, the difference is relatively small (about a factor of two) giving us confidence in the more involved discovery limit calculation presented in the main body of the paper.

\section{Comparing neutron-only and proton-only spin-dependent constraints}
\label{app:proton}

In the main body of the paper we only compared the discovery reach against the exclusion limits from two-phase xenon experiments, which are sensitive to the spin-dependent interaction with the neutron. In this appendix we discuss the limits from experiments that are sensitive to the spin-dependent interaction with the proton. A comparison of the neutron-only and proton-only spin-dependent limits are necessarily model dependent. The cross-sections from the axial-vector interaction are related by
\begin{equation}
\frac{\sigma_p^0}{\sigma_n^0}=\left(\frac{\sum_{q=u,d,s} \,g_q\, \Delta_q^p}{\sum_{q=u,d,s}\, g_q\, \Delta_q^n}\right)^2\;,
\end{equation}
where the $\Delta$ factors encode the fraction of the nucleon spin carried by the labelled quark in a proton or neutron. We use the values recommended by the PDG~\cite{Agashe:2014kda}
\begin{equation}
\Delta^p_u=\Delta^n_d=0.84\,,\qquad \Delta^p_d=\Delta^n_u=-0.43\,,\qquad \Delta^p_s=\Delta^n_s=-0.09\;.
\end{equation}
The factors~$g_q$ are the coupling constants of the axial-vector mediating particle with the quarks and come from the dark matter model. When~$g_q$ is the same for all quarks, $\sigma_p^0=\sigma_n^0$ so the limits can be trivially compared. Arguably a better motivated choice is $g_u=-g_d=-g_s$. This is the case with the $Z$-boson in the Standard Model and is therefore applicable to well-motivated dark matter candidates such as the neutralino. In this scenario, we find that $\sigma_n^0=0.75\times\sigma_p^0$.

\begin{figure}[t!]
\centering
\includegraphics[width=0.49\columnwidth]{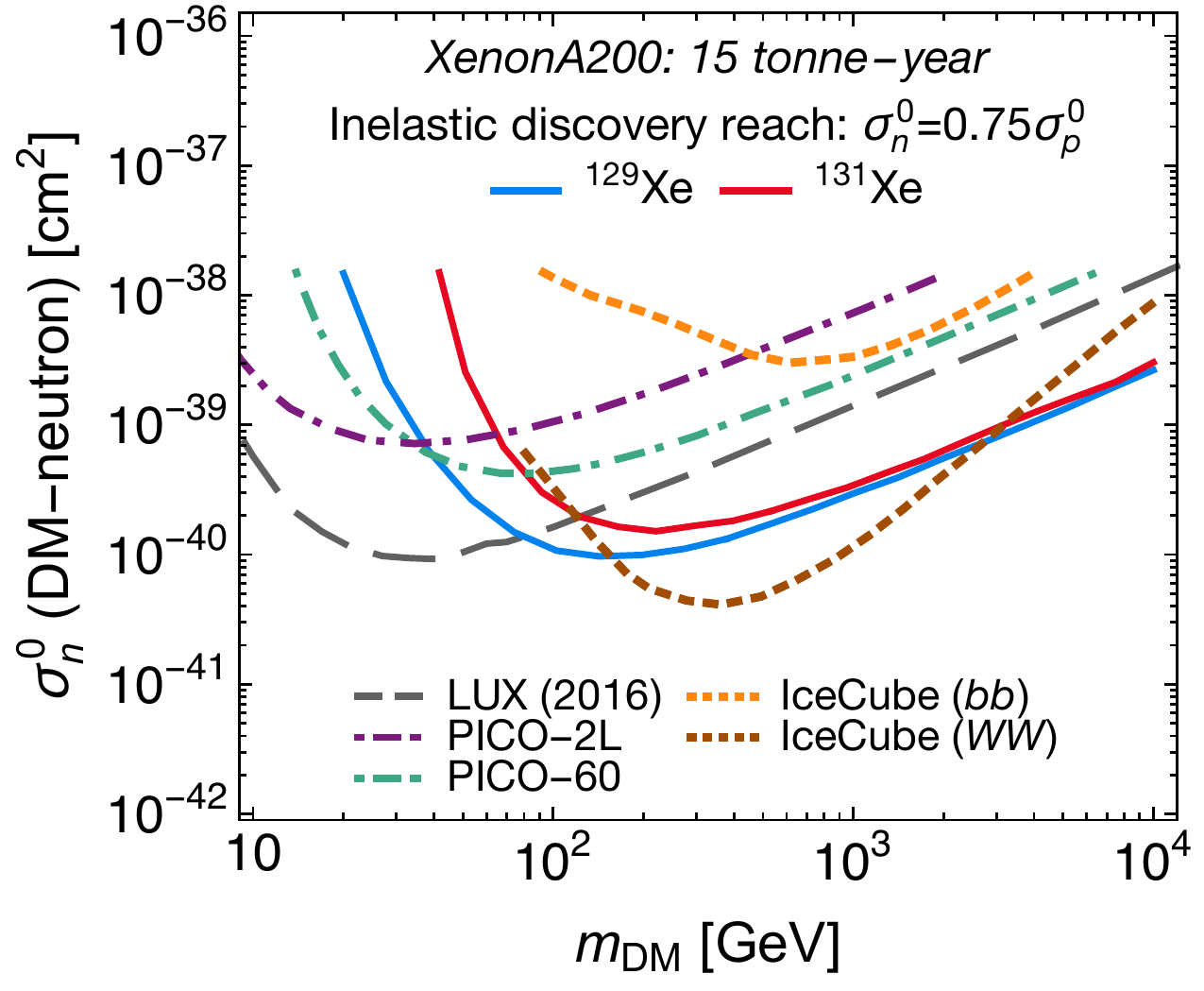} 
\includegraphics[width=0.49\columnwidth]{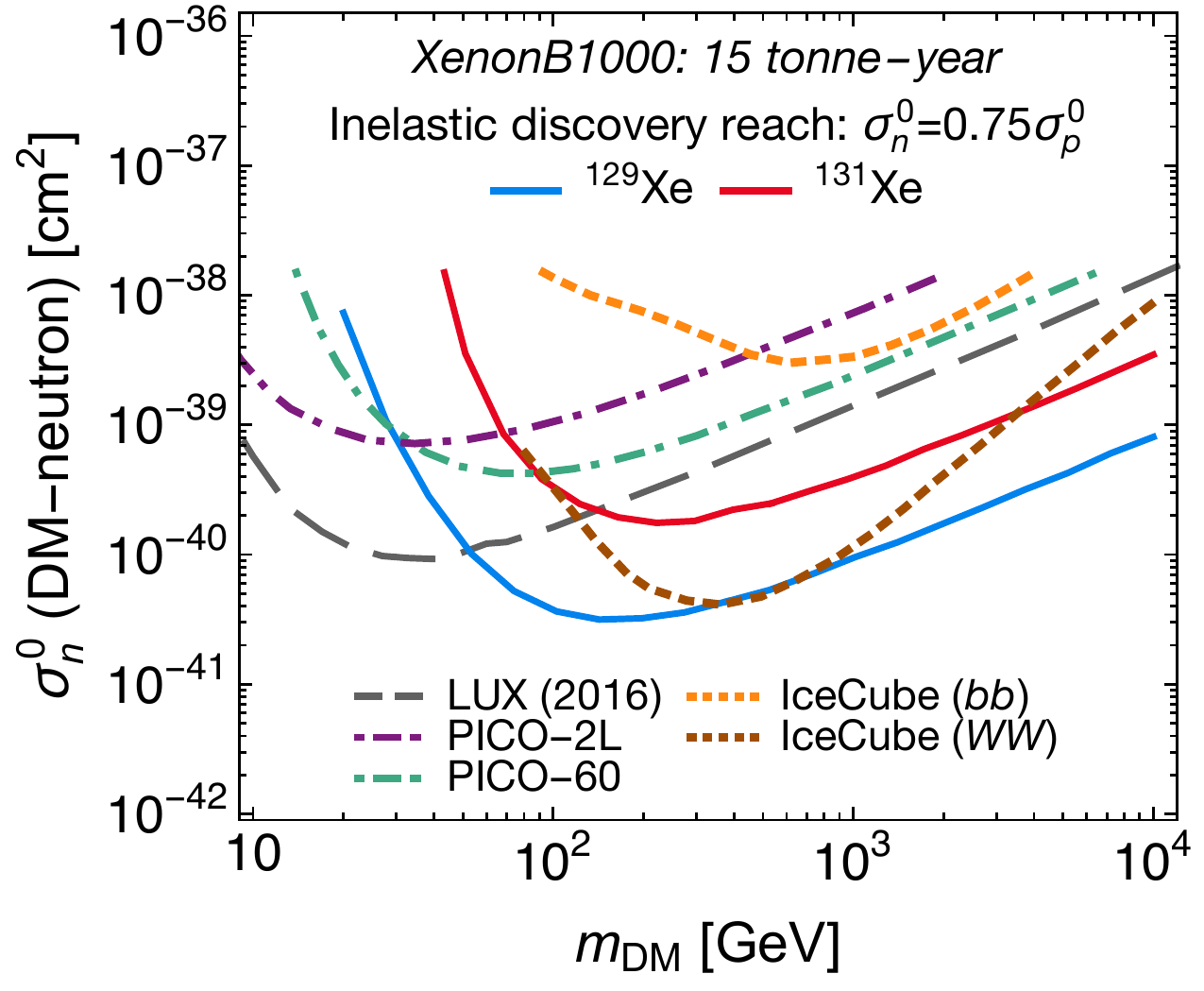}
 \caption{A comparison of the proton-only spin-dependent limits from PICO-2L, PICO-60 and IceCube with the exclusions limit from LUX and the discovery reach for inelastic scattering off the~$\xea$ and~$\xeb$ isotopes. We have assumed that $\sigma_n^0=0.75\times\sigma_p^0$, the relationship that holds, for instance, with the $Z$-boson in the Standard Model, and rescaled the published PICO and IceCube limits appropriately. The IceCube limits are comparable to the discovery reach over part of the mass range but are rather model-dependent (see text for details) and it is straightforward to envisage scenarios in which these limits do not apply. The LUX limit is model-independent and is always below the PICO limits, justifying our choice to show only the LUX limits in the main body of the paper.  \label{fig:plimit}}
 \end{figure}

The strongest direct detection constraints on the proton-only spin dependent cross-section are from PICO2L~\cite{Amole:2015lsj} and PICO-60~\cite{Amole:2015pla} experiments. These are plotted as the dot-dashed green and purple lines in figure~\ref{fig:plimit}. In this figure, we have assumed that $\sigma_n^0=0.75\times\sigma_p^0$ and rescaled the published PICO limits appropriately. By comparing with the LUX limit (grey long-dashed line) we observe that the LUX limit is always below the PICO limits. This justifies our choice of only showing the LUX limits in the main body of the paper since this experiment is the most constraining direct detection experiment for this type of interaction.

There are also constraints on the proton-only spin-dependent cross-section from the IceCube search for neutrinos originating from the Sun~\cite{Aartsen:2016exj}. These limits are valid when the dark matter capture and annihilation rates are in equilibrium (which must be checked on a model-by-model basis). The limits also depend on the dark matter annihilation final-state. The strongest limits are for annihilation to $W^+ W^-$ (brown-dashed line in figure~\ref{fig:plimit}), weaker limits are obtained for annihilation to $\bar{b}b$ (orange-dashed line in figure~\ref{fig:plimit}), while no limits exist for annihilation to first and second generation quarks and leptons. As with the PICO limits, we have assumed that $\sigma_n^0=0.75\times\sigma_p^0$ and rescaled the published limits appropriately. In this instance, we see that the IceCube $W^+ W^-$ limit is comparable to the discovery reach from a 15~tonne-year two-phase xenon experiment for a dark matter mass below $\sim1000$~GeV. However, given the model-dependent assumptions that enter the IceCube constraints, it is possible to consider scenarios in which the IceCube limit does not apply. In comparison, the LUX limit is model independent.

\section{The discovery reach with a halo model from EAGLE}
\label{app:Halo}

\begin{figure}[t!]
\centering
\includegraphics[width=0.54\columnwidth]{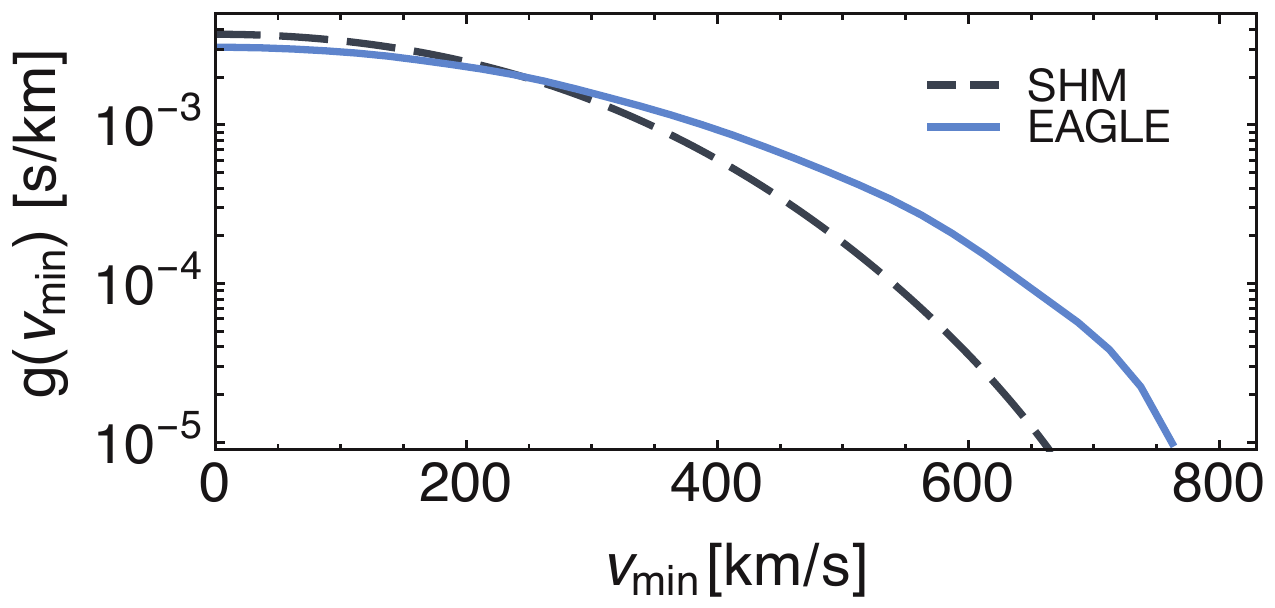} 
 \caption{A comparison of the velocity integral $g(v_{\rm{min}})$ for two halo models: the Standard Halo Model (SHM), the model we used in the main body of the paper, and the Milky Way-like halo that deviates most from the SHM from the EAGLE HR simulation. The EAGLE velocity integral contains more particles with a higher velocity.
 \label{fig:halocomp}}
 \end{figure}
 
  \begin{figure}[t!]
\centering
\includegraphics[width=0.49\columnwidth]{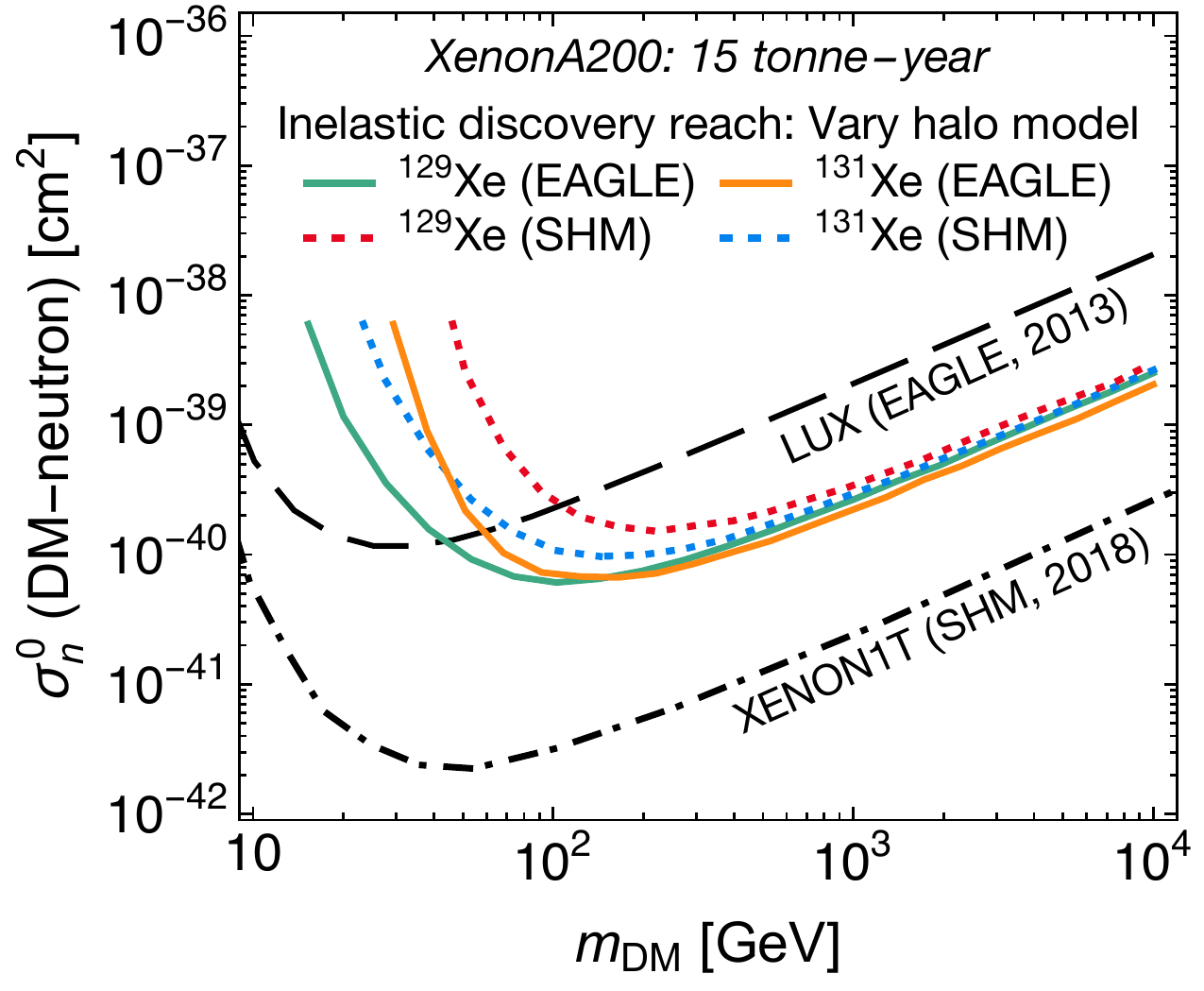} 
\includegraphics[width=0.49\columnwidth]{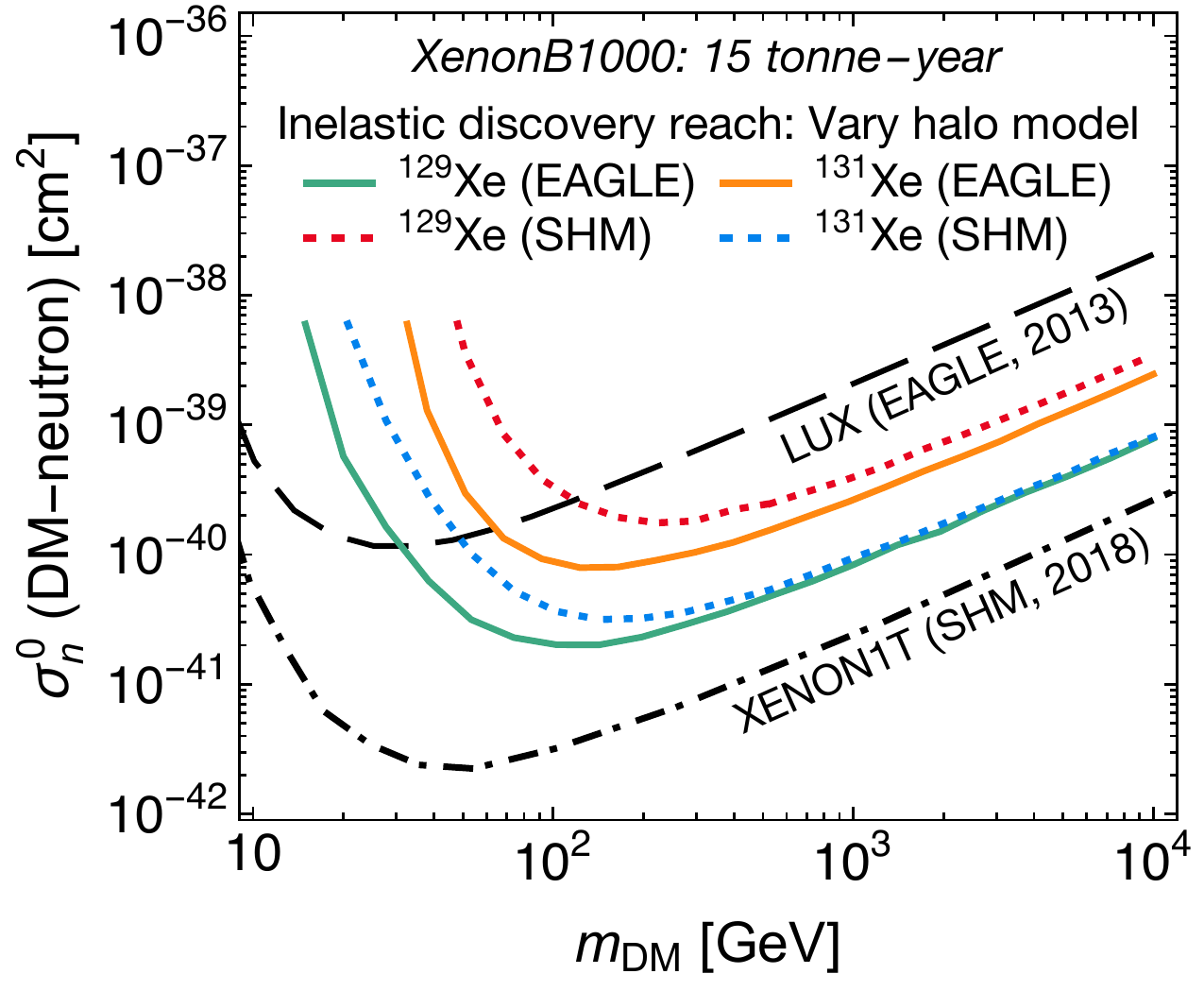}
 \caption{A comparison of the discovery reach assuming the Standard Halo Model (dashed red and blue lines) and the EAGLE halo model (solid green and orange lines) for inelastically scattering off the~$\xea$ and~$\xeb$ isotopes. The left and right panels show the results for the {\it XenonA200} and {\it XenonB1000} benchmark scenarios assuming a 15~tonne-year exposure. Also shown in the LUX exclusion limit (black dashed line) assuming the EAGLE halo model and the XENON1T projected limit assuming the SHM. Although the discovery reach assuming the SHM and EAGLE halo models show differences at smaller values of the dark matter mass, the conclusions drawn from this figure are essentially the same as in figure~\ref{fig:limit}, showing that our overall results are not particularly sensitive to the choice of halo model.  \label{fig:halolimit}}
 \end{figure}
 
We assumed the Standard Halo Model (SHM) in all of the calculations in the main body of the paper. This is the simplest and canonical halo model used by the direct detection community. In this appendix we investigate how our results change when another halo model is used. It is well known that the choice of halo can result in large deviations in the scattering rate for scattering processes that involve an inelastic transition. This is because these processes typically probe dark matter particles that are towards the tail of the velocity distribution~\cite{MarchRussell:2008dy,McCabe:2010zh}. The alternative halo that we consider in this appendix is from the EAGLE simulation~\cite{Schaye:2014tpa,Crain:2015poa}, which is a state-of-the-art simulation of galaxies that contains the effects of dark matter and baryons. The simulation contains a vast number of halos so various criteria must be used to select those that are Milky Way-like. This procedure was implemented in~\cite{Calore:2015oya, Bozorgnia:2016ogo}. The quantity which enters the event rate is the velocity integral
\begin{equation}
g(v_{\rm{min}})=\int_{v_{\rm{min}}}\frac{f(v)}{v}\; d^3v\;.
\end{equation}
Here we use the velocity integral that deviates most from the SHM while passing all of the selection criteria from the EAGLE~HR simulation run in~\cite{Bozorgnia:2016ogo}. This is shown as the blue solid line in figure~\ref{fig:halocomp}, where it is compared with the halo integral from the Standard Halo Model, the dashed black line. The main difference is that the EAGLE halo contains many more particles at higher values of $v_{\rm{min}}$.

In figure~\ref{fig:halolimit} we show the discovery reach when using the EAGLE halo model and compared with the discovery reach from the SHM. As we would naively expect based on~\cite{MarchRussell:2008dy,McCabe:2010zh}, we find that the largest difference between the halo models is at low mass and when the inelastic splitting is large. This is simply because the EAGLE halo has more particles with a higher speed, so an experiment observes more events when the dark matter mass is below approximately 100~GeV. We have also recalculated the LUX exclusion limit from their 2013 search~\cite{Akerib:2013tjd} for dark matter that elastically scatters with the xenon isotopes with the EAGLE halo (we were not able to recalculate the XENON1T projected limit since we do not know all of the assumptions entering its calculation). The calculation of this limit is described in~\cite{Buchmueller:2014yoa}. Overall, we see that our conclusions remain essentially unchanged; the discovery reach still lies between the LUX limit and XENON1T projected limit. The small difference is the that the SHM results in a slightly more conservative discovery reach, with the EAGLE discovery reach extending to slightly smaller values of the dark matter mass.

\bibliography{ref}
\bibliographystyle{JHEP}

\end{document}